\documentclass[fleqn,10pt]{wlscirep}
\usepackage[utf8]{inputenc}
\usepackage[T1]{fontenc}
\usepackage{booktabs}
\usepackage{tabularx}
\usepackage{array}
\usepackage{graphicx}
\usepackage{multirow}  
\usepackage{colortbl}
\definecolor{gainlow}{RGB}{242,247,252}
\definecolor{gainmid}{RGB}{217,232,245}
\definecolor{gainhigh}{RGB}{188,214,236}
\definecolor{gainvhigh}{RGB}{158,195,224}
\usepackage[table]{xcolor}
\usepackage{array}
\definecolor{consortHead}{HTML}{B8CCE4}
\definecolor{consortPanelA}{HTML}{E9EBEE}
\definecolor{consortPanelB}{HTML}{DCDFE3}
\definecolor{consortText}{HTML}{2B2B2B}
\newcounter{methodssection}

\title{Bridging GenAI policy and practice: Brief instruction improves learning-oriented prompting in an introductory programming course}

\author[1,*]{Ruiwei Xiao}
\author[2]{Runlong Ye}
\author[3]{Xinying Hou}
\author[4]{Jessica Wen}
\author[2]{Harsh Kumar}
\author[4]{Michael Liut}
\author[1]{John Stamper}

\affil[1]{Carnegie Mellon University, Pittsburgh, PA, USA}
\affil[2]{University of Toronto, Toronto, ON, Canada}
\affil[3]{New Jersey Institute of Technology, Newark, NJ, USA}
\affil[4]{University of Toronto Mississauga, Mississauga, ON, Canada}

\affil[*]{ruiweix@cs.cmu.edu}

\keywords{generative AI, learning-oriented prompting, randomized controlled trial (RCT), cognitive engagement, higher education, AI literacy, AI in education}

\begin{abstract}
Generative artificial intelligence (GenAI) is now commonplace in higher education, yet unguided use can harm learning. Universities have responded with policies asking students to use GenAI for learning, not solutions, but how to translate these policies into classroom instruction remains untested. Here we report a randomized controlled trial with a six-week, end-of-semester follow-up ($N = 431$) in an introductory programming course comparing four instructional approaches, from a two-sentence reminder (Condition~1) to a multi-step constructive task (Condition~4). We found: (1) all four approaches produced learning-oriented prompting gains that persisted six weeks later and scaled with the designed engagement level, from 8 to 51 percentage points; (2) students in every condition self-reported sustained voluntary use, and confidence grew in step with skill; and (3) final examination performance did not differ between conditions, placing the possible boundary of the effects at the pathway from prompting skill to achievement. These findings provide instructors with an empirically calibrated set of options for translating GenAI-use policies into classroom practice.
\end{abstract}
\begin{document}

\flushbottom
\maketitle
\thispagestyle{empty}

\section*{Introduction}

Generative artificial intelligence (GenAI) has become part of everyday academic life for university students, with 94\% of undergraduate students reporting use for academic work\cite{hepi2026} and four in five university students worldwide using GenAI for coursework by 2026\cite{stanfordhai2026}. The rapid integration of GenAI has not only changed the process by which students study\cite{frydenberg2026rapid, fang2026research}, write \cite{gabay2026generative}, and solve problems \cite{hou2026effects}, but also changed the cognitive demands they actually engage with\cite{yan2024promises,kasneci2023, li2026cognitive, zhang2026cognitive}. A growing body of evidence suggests that use of GenAI may encourage cognitive offloading \cite{gerlich2025ai}, and that over-reliance on AI-generated content without critical evaluation is linked to shallower engagement and reduced critical thinking \cite{suazo2026between, lee2025,fan2025metacognitive}. Empirical studies also show that students' learning can be undermined when they rely on GenAI\cite{darvishi2024impact, stadler2024cognitive,barcaui2025chatgpt}. For example, a recent randomized trial found that students given access to an unrestricted LLM tutor performed 17\% worse on subsequent unassisted assessments than peers without access in math \cite{bastani2025}. A key concern about students' use of AI for learning is that students primarily use AI to obtain immediate answers rather than to engage in deeper, learning-oriented inquiry, potentially limiting active engagement, independent thinking, and meaningful learning \cite{yan2025beyond,iqbal2025}.

In response, universities and instructors have increasingly adopted policies encouraging students to use GenAI ``for learning, not solutions'' \cite{jin2025,sullivan2023,kestin2025}. However, these policies often emphasize broad institutional principles \cite{jiang2025exploring} or classroom-level rules \cite{wu2026governing,mcdonald2025generative}, offering limited practical guidance on how students can formulate learning-oriented queries when interacting with AI chatbots. Although recent research has introduced frameworks to support more regulated GenAI use \cite{wang2026idea,lin2026designing}, AI literacy instruction has mostly taught students to prompt for better outputs rather than for learning. Existing classroom interventions have improved students' prompting skills and AI self-efficacy \cite{theophilou2023learning,woo2026effects}, but they targeted the quality of AI responses, not whether those responses support learning. In programming education, where our study is situated, prompting instruction has largely focused on getting AI to generate correct code \cite{denny2024prompt}, while unguided use can widen the gap between stronger and struggling novices \cite{prather2024widening}.

Concrete instructional guidance for translating these principles into effective learning-oriented queries therefore remains limited, leaving three practical questions underexplored. First, what specific prompting strategies should students adopt to elicit learning-oriented interaction with AI \cite{xiao2025,knoth2024}? Second, how should instructors teach these strategies given limited classroom time? Third, does such instruction actually shift students' prompting behavior, and does any shift persist? In the absence of evidence, instructors face a structural mismatch: students have access to powerful generative systems but lack concrete guidance on using them in ways that support, rather than substitute for, their own thinking \cite{ashraf2025,fan2025educational}.

Prior literature has identified strategies for prompting AI to behave as a tutor rather than merely as a source of direct answers~\cite{xiao2025,knoth2024,lee2025prompt, kim2026prompting}, and has reported preliminary evaluations of their effectiveness. Here we adopt the design from Xiao et al.\cite{xiao2025} and refer to the target competency as \textit{learning-oriented prompting}: the construction of queries that elicit explanation, scaffolding, and verification rather than direct answers\cite{xiao2025,white2023,zamfirescu2023}. Building on previous literature, we adopt the six-component operationalization of learning-oriented prompting proposed by Xiao et al.\cite{xiao2025}: five learning-context components, namely the AI's role (the tutoring persona the AI should adopt), the learner's level (calibrating output complexity to the learner's proficiency), problem context (providing grounding information such as code or error messages), difficulty identification (articulating one's knowledge gap), and guardrails (explicit constraints on what the AI should not provide), together with a sixth component, the tutoring protocol (specifying a pedagogical strategy such as worked examples or scaffolded hints). This six-component framework provides an evidence-based, actionable basis for translating learning-oriented GenAI-use policies into classroom instruction. Accordingly, all prompting outcomes in this study are operationalized as alignment with this six-component framework. The empirical question is whether brief instruction in this framework can durably shift how students prompt AI for learning.

To implement instruction and explore how to teach effectively, we draw on the Interactive--Constructive--Active--Passive (ICAP) framework\cite{chi2014} as a tool for varying the engagement and effort the instruction demands. ICAP proposes that learning gains scale with the depth of overt cognitive engagement: passive reception produces the smallest gains; active manipulation produces larger gains; constructive generation produces larger gains still. ICAP has accumulated substantial empirical support in well-defined STEM domains\cite{freeman2014}, providing a principled basis for designing a graded sequence of instructional approaches. We draw on this robust gradient not to test whether ICAP holds, but to address a practical question it leaves open: because tools that package well-constructed prompts already exist, instructors and students need a reason to invest time in learning to construct prompts, and evidence on how much engagement such instruction must demand to be worthwhile. Using ICAP to define a graded sequence of approaches allows us to estimate the benefit and time cost of each level in an authentic classroom. Three research questions guided the investigation:

\vspace{-1mm}
\begin{description}
    \setlength{\itemsep}{-1pt}
    \setlength{\parsep}{-1pt}
    \setlength{\topsep}{-1pt}
    \item[\textbf{RQ1.}] Do different instructional approaches durably improve students' learning-oriented prompting?
    \item[\textbf{RQ2.}] Do these approaches have a lasting influence on how students subsequently learn with GenAI?
    \item[\textbf{RQ3.}] Do the effects vary across measured learner characteristics?
\end{description}
\vspace{-3mm}

\section*{Methods}
\phantomsection
\refstepcounter{methodssection}
\label{sec:methods}

\subsection*{Participants}

Participants were first-year undergraduate students enrolled in CSC108, an introductory computer science course at the University of Toronto Mississauga, Canada, in fall 2025. All 1024 students enrolled in the course were randomly assigned to one of four conditions by a course-management script using a uniform pseudorandom generator with no blocking ($n = 251$, $255$, $254$, and $264$ in Conditions~1--4). The analytic sample comprised the 431 students who completed all four required study phases (pre-test, intervention, immediate post-test, and delayed post-test; $n = 102$, $112$, $109$, and $108$; Supplementary Fig.~S1).

In the analytic sample, 82\% of participants who reported their program of study ($n = 385$) were in computer science; of those who reported gender ($n = 328$), 70\% identified as men, 29\% as women, and 1\% as non-binary; and 34\% of those who reported student status ($n = 385$) were international students (Supplementary Table S1). Age was not collected under the approved ethics protocol; all participants were adults (aged 18 or older), a population typically aged 18 to 22 given the first-year composition of the course. Pre-test scores did not differ significantly between conditions, suggesting that randomization produced groups comparable in baseline prompting ability, nor did any other measured baseline characteristic (Supplementary Table S1).

Attrition before the Week-1 pre-test (1024 to 534) reflects enrollment churn during the university's add/drop period, students who declined consent for research use of their data, and incompletion of the survey; all pre-test completers also completed the intervention and immediate post-test, which were mandatory course activities; attrition before the delayed post-test (534 to 431) is consistent with its administration as an optional extra-credit item on the final examination, although individual reasons for non-completion (for example, insufficient remaining time during the examination) were not recorded. Attrition did not differ significantly between conditions at any stage (allocation to pre-test: $\chi^2(3) = 1.10$, $p = 0.777$; immediate to delayed post-test: $\chi^2(3) = 0.99$, $p = 0.805$) or overall (allocation to analytic sample: $\chi^2(3) = 0.79$, $p = 0.852$), and students who did not complete the study did not differ from completers in baseline prompting scores (Welch $t = 0.40$, $p = 0.692$), indicating no evidence of condition-dependent or ability-dependent attrition. Sixteen students in Conditions~3 and~4 completed all study phases but recorded no valid interaction with their assigned task; they were retained in the analytic sample, consistent with its definition, and a sensitivity analysis excluding them ($n = 415$) reproduced all between-condition effects with an identical pattern of pairwise differences ($F(3, 411) = 70.75$, $p < 0.001$).

\subsection*{Study design and ethics}

\begin{figure}[!ht]
\centering
\includegraphics[width=\textwidth]{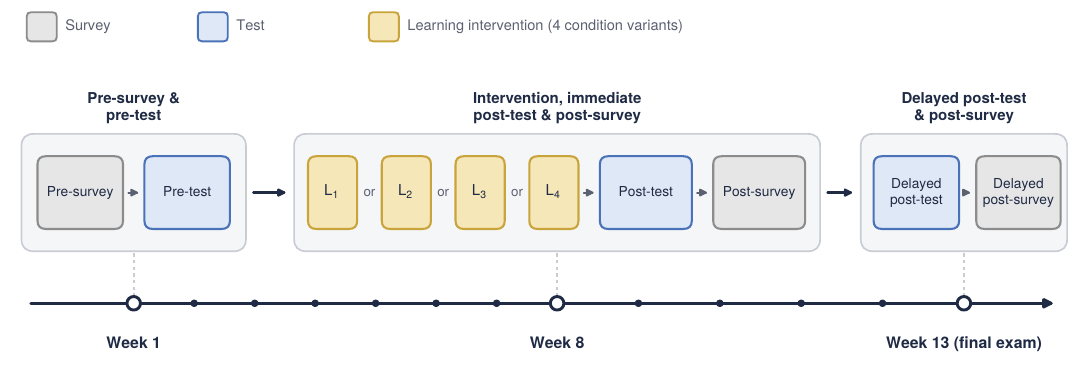}
\caption{Study timeline. Students completed a pre-survey and pre-test in Week~1; were randomly assigned to one of four instructional conditions in Week~8 (L1: brief passive reminder; L2: scenario-based reading with worked examples; L3: component selection; L4: select-then-write with automated feedback) and completed an immediate post-test and post-survey within the same take-home lab assignment; and completed a delayed post-test and delayed post-survey as an optional bonus question in the final examination in Week~13.}
\label{fig:design}
\end{figure}

We conducted a semester-long randomized controlled trial within a mandatory introductory computer science course at a large public North American university (Fig.~\ref{fig:design}). The protocol was approved by the institutional research ethics board, and all participants provided informed consent. Students completed a pre-survey and pre-test at the start of the semester (Week~1); were randomly assigned to one of four instructional conditions (L1--L4) during a mandatory ``Ethical Use of AI for Learning Lab'' in Week~8, immediately followed by an immediate post-test and post-survey; and completed a delayed post-test as an extra-credit item on the final examination at the end of Week~13, approximately six weeks after the intervention. The pre-survey, the immediate post-survey, the pre-test and immediate post-test, and the intervention itself were administered online as take-home activities that students completed on their own devices, with access to generative AI chatbots. The delayed post-test and delayed post-survey were the exception: they were completed with paper and pencil during the proctored, in-person final examination, without access to any generative AI tool.

\subsection*{Study materials}

\subsubsection*{Intervention design}

\begin{figure}[!ht]
\centering
\includegraphics[width=\textwidth]{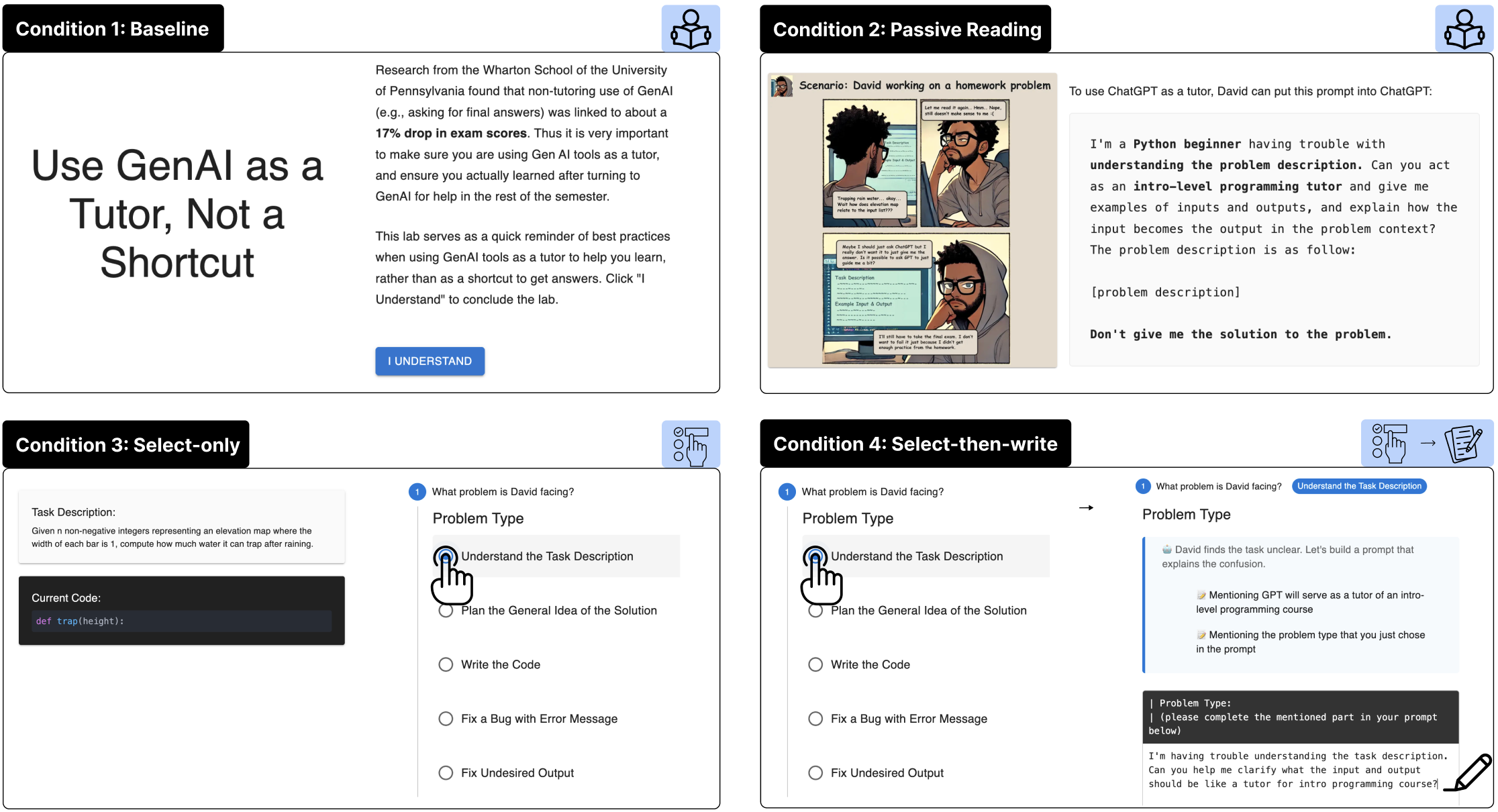}
\caption{The four instructional conditions, occupying ascending levels of cognitive engagement under the ICAP framework. Condition~1 (baseline): a two-sentence textual reminder framed by recent evidence on the cost of unguided GenAI use. Condition~2 (passive reading): scenario-based comics presenting a fictional peer applying an appropriately constructed pedagogical prompt. Condition~3 (active selection): students select, for each rubric component, the most appropriate option from three alternatives. Condition~4 (select-then-write): students additionally compose the full prompt; an automated system validates each component against pre-specified acceptance criteria and provides actionable feedback. Conditions~2--4 presented the same cartoon and the same three scenarios and differed only in the form of interaction; the figure shows only the elements that differed between conditions. Full screenshots are provided in Supplementary Figs.~S2--S5.}
\label{fig:conditions}
\end{figure}

The four conditions were designed \textit{a priori} to occupy ascending levels of the ICAP framework\cite{chi2014}, specifying a lower bound on cognitive engagement without constraining the upper bound (Fig.~\ref{fig:conditions}).

\textbf{Condition~1 (brief passive reminder)} presented a two-sentence textual reminder instructing students to use GenAI as a tutor rather than a shortcut, framed by recent evidence that unguided GenAI use can impair learning\cite{bastani2025}. This condition mirrors the brief advisory guidance institutions commonly provide. Students in this condition were only required to read the reminder; no response was collected.

\textbf{Condition~2 (passive worked example)} presented three scenario-based comic narratives in which a fictional peer encountered each problem type defined by the framework and applied an appropriately constructed pedagogical prompt. Students in this condition were required to read each help-seeking scenario, set in an introductory programming course, together with the corresponding example pedagogical prompt that steers the AI to act as a tutor; as in Condition~1, no response was collected.

\textbf{Condition~3 (active selection)} required students to analyze the same scenarios but to select, for each rubric component, the most appropriate option from three alternatives. The correctness of each selection was checked against pre-specified answers by a rule-based script. Students in this condition were required to read each scenario and assemble a pedagogical prompt step by step, selecting one of three options for each of the six essential components, repeating this on three isomorphic scenarios.

\textbf{Condition~4 (constructive practice)} extended Condition~3 by additionally requiring students to compose the full prompt; an automated system based on GPT-4o validated each component against pre-specified acceptance criteria and provided actionable feedback when components were missing or incomplete, supporting iterative refinement. Students in this condition were required to read each scenario and assemble the prompt one component at a time: for each of the six components, they first selected the most appropriate option, as in Condition~3, and then immediately wrote that component in their own words, revising it in response to the automated feedback; the six written components together formed the complete prompt.

The scenarios were identical across Conditions~2--4; the scenarios and example prompts used in the intervention were distinct from, although isomorphic in structure to, those used in the pre- and post-tests. We did not collect a direct manipulation check on cognitive engagement; we discuss this limitation in Limitations and future work. Time-on-task was logged by the study platform; median durations by condition are reported in Table~\ref{tab:learning_gains}.

\begin{figure*}[t]
  \centering
  \includegraphics[width=.78\textwidth]{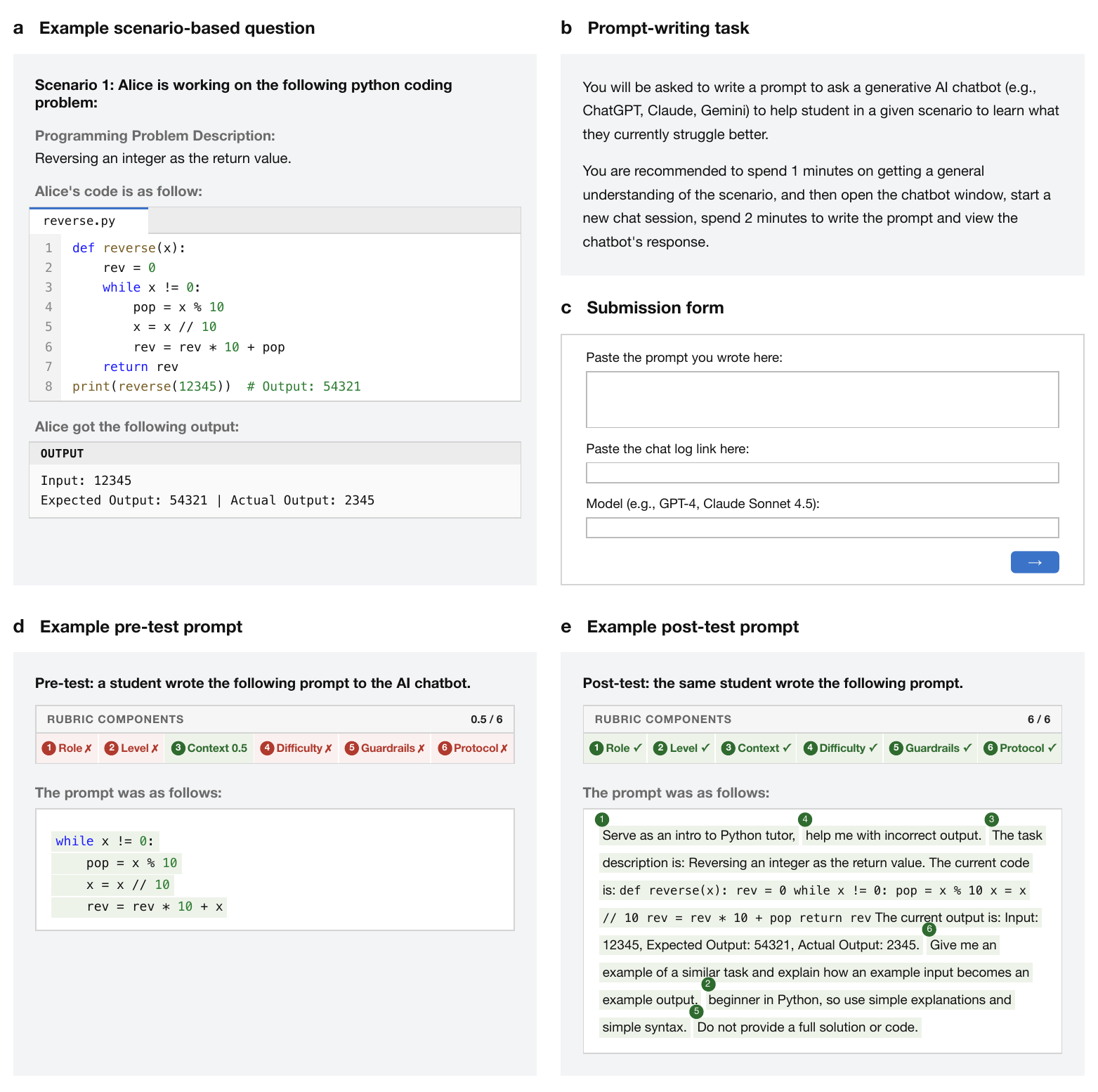}
  \caption{Study task and example prompts.
  \textbf{(a)}~An example scenario question: a student's Python code for reversing an integer gives the wrong output.
  \textbf{(b)}~Task instructions: participants write a prompt asking an AI chatbot to help the student learn.
  \textbf{(c)}~The submission form.
  \textbf{(d--e)}~Prompts the same participant wrote before (d) and after (e) the intervention, scored on six rubric components: \textit{Role}, \textit{Level}, \textit{Context}, \textit{Difficulty}, \textit{Guardrails} and \textit{Protocol}. Green highlights mark the text that meets each component. The pre-test prompt scores 0.5/6 and the post-test prompt scores 6/6.}
  \label{fig:task_prompts}
\end{figure*}

\subsubsection*{Pre- and post-test}

The primary outcomes were learning-oriented prompting scores at three time points: pre-test, immediate post-test (administered within the same take-home lab assignment as the intervention), and delayed post-test (administered approximately six weeks later, in paper-and-pencil format, at the end of the final examination). The pre-test and immediate post-test each comprised three prompt-writing questions; the delayed post-test, administered within the final examination, comprised a single prompt-writing question. Each question presented a programming scenario in which a fictional peer encountered a specific coding obstacle, and asked the student to write a prompt to a GenAI tool that would help the peer learn rather than simply receive the answer (Fig. \ref{fig:task_prompts}a--c). All questions were isomorphic in structure; pre- and immediate post-test questions targeted concepts recently taught in CSC108 (e.g., while loops), while the delayed post-test question targeted database foreign keys, an advanced topic not covered in the course. Each student’s score at each time point was computed as the mean across the three questions; the delayed post-test score was the score on its single question.

\subsubsection*{Pre- and post-survey}

The pre-survey collected demographic and academic background items alongside three validated instruments. Gender, program of study, and international student status were collected through self-report items. For the exploratory moderation analysis, gender was represented as women and non-binary students versus men because the non-binary subgroup was too small for a separate inferential comparison. A sensitivity analysis restricted to women versus men yielded a similar directional pattern. Implicit theories of intelligence were measured with an eight-item scale\cite{dweck2013} on a 6-point Likert response (four entity items, four incremental items, the former reverse-scored to yield a composite where higher values indicate stronger growth mindset). Need for cognition was measured with the 18-item short form\cite{cacioppo1984}. Self-perceived programming ability was measured with items adapted from a validated computer science self-efficacy instrument\cite{tew2010}, yielding two subscales: concept understanding and programming self-efficacy.

The immediate post-survey, completed directly after the immediate post-test, measured confidence in prompting with two items rated on a 5-point scale: ``Now I know how to effectively use LLM AI tools like ChatGPT to support my learning'' and, retrospectively, ``Before this experiment, I know how to effectively use LLM AI tools like ChatGPT to support my learning.'' Confidence gain was computed as the difference between the two ratings.

The delayed post-survey, completed in paper-and-pencil format alongside the delayed post-test on the final examination, measured voluntary use of pedagogical prompting with a single item: ``Since completing Lab 7.5, how many times have you prompted a generative AI tool (e.g., ChatGPT, Gemini) for pedagogical/learning purposes (e.g., asking to explain step-by-step, without revealing the solution)?'' Lab 7.5 was the Week-8 lab in which the intervention was delivered. Responses were coded from 1 (never) to 5 (every day), with intermediate options of once a month, once a week, and more than once a week. The item asked about use in any context rather than within CSC108 specifically. Single-item measures were used throughout to minimize survey burden in this classroom setting; single-item self-reports show adequate validity for narrow, concrete constructs of this kind\cite{wanous1997overall}.

\subsection*{Qualitative coding}

\subsubsection*{Prompt-writing items }Each prompt was scored against the six-component rubric (AI's role, learner's level, problem context, difficulty identification, guardrails, and tutoring protocol; described in the Introduction). Each component was scored on a 0/0.5/1 scale, yielding a maximum of 6 points per prompt; a prompt's score is reported as the percentage of this maximum. Figure~\ref{fig:task_prompts}d--e shows a weak and a strong participant-written prompt with the corresponding component-level scores. Three trained research assistants first independently coded a stratified random sample of 10\% of the prompts; inter-rater reliability was Fleiss' $\kappa = 0.86$. After resolving the disagreements, the remaining prompts were divided among the assistants for independent coding. Handwritten responses from the delayed post-test were transcribed and scored with the same rubric. The rubric was never shown to students at any phase of the study. The full rubric is provided in Supplementary Table~S3.

\subsubsection*{Open-ended post-survey item }The delayed post-survey asked students to \textit{``write one prompt that you personally have used that was helpful for your learning, and explain briefly why it was helpful to you.''} As a supplement to RQ2, one coder labeled the subject or course context described in each response, and the frequency of contexts outside CSC108 is reported descriptively.

\subsection*{Statistical analysis}

All analyses were conducted in Python (version~3.11). Within-condition learning gains were tested using two-sided paired $t$-tests, or Wilcoxon signed-rank tests where the normality assumption was violated (Shapiro--Wilk $p < .05$); paired $t$-tests compared pre-test with post-test scores, so negative $t$ values correspond to positive gains. Between-condition comparisons of learning gains, voluntary use, confidence gains, and final examination scores used one-way analysis of variance with Tukey HSD post-hoc tests; effect sizes for these comparisons are reported as $\eta^2$. Effect sizes for within-condition gains are reported as Cohen's $d$, computed as the mean gain divided by the standard deviation of gains.

For RQ1, component-level gains were computed for each rubric component as the per-student change from pre-test to immediate post-test, averaged within condition. As a sensitivity analysis, we re-estimated between-condition differences in learning gains while controlling for log-transformed time-on-task, winsorized at the within-condition 95th percentile to remove idle-session outliers; medians of time-on-task are reported with bootstrap 95\% confidence intervals (10,000 resamples).

For RQ2, the association between individual confidence gains and prompting skill gains was tested with Spearman rank correlation. Final examination scores were regressed on confidence gains (OLS), controlling for condition, prompting skill gains, and pre-test scores; a second model, controlling for condition and pre-test scores, tested the condition-by-confidence-gain interaction jointly with an $F$ test. Overconfidence was indexed as each student's standardized confidence gain minus standardized prompting skill gain, and its association with examination scores was tested with Spearman rank correlation. Equivalence of final examination scores between conditions was assessed with two one-sided Welch tests (TOST) per condition pair, using equivalence bounds of $\pm 5$ percentage points, supplemented by 90\% confidence intervals for each pairwise difference.

For RQ3, exploratory moderation analyses used ordinary-least-squares models predicting immediate and delayed gains from condition, each candidate moderator, and their interaction. Unadjusted $p$ values are reported, and interactions that did not survive multiple-testing correction are treated as hypothesis-generating. The gender interaction was additionally examined using permutation testing and a baseline-adjusted sensitivity model. All tests are two-sided with $\alpha = .05$, except the one-sided tests comprising the TOST procedure.

\section*{Results}

\subsection*{RQ1: All instructional approaches durably improved learning-oriented prompting, with learning gains increasing with the designed engagement level}

To answer RQ1, we examined learning-oriented prompting scores within each condition, immediately and at six-week follow-up, and compared the magnitude and durability of gains between conditions.

\textit{All approaches, even the briefest, produced durable gains.} All four conditions produced statistically significant gains in learning-oriented prompting at the immediate post-test, with mean gains ranging from $8$ percentage points ($\mathit{SD}=15$) in the brief passive reminder condition (Condition~1) to $51$ ($\mathit{SD}=28$) in select-then-write (Condition~4) (Table~\ref{tab:learning_gains}; all $p < .001$). At the delayed post-test on week 13, conducted as part of the final examination, all four within-condition gains remained statistically significant (mean delayed gains $6$--$24$ percentage points; all $p < .001$). The persistence of gains in the brief passive condition (Condition~1) is particularly notable: students who read only a two-sentence reminder reframing GenAI as a tutor rather than an answer source produced prompts whose structure remained significantly altered six weeks later at the end of the semester, in the absence of booster sessions or in-course reinforcement (paired $t(101)=-4.00$, $p<.001$, $d=0.38$).

\begin{table}[!ht]
\centering
\small
\setlength{\tabcolsep}{4pt}
\vspace{-2mm}
\caption{\textbf{Learning-oriented prompting scores and learning gains at the immediate and six-week delayed post-tests.}}
\vspace{-2mm}
\label{tab:learning_gains}
\begin{tabular}{llcccccccc}
\toprule
Condition & \textit{n} & \begin{tabular}[c]{@{}c@{}}Pre-test, \%\\mean (s.d.)\end{tabular} & \begin{tabular}[c]{@{}c@{}}Duration, min\\median [95\% CI]\end{tabular} & Time point & \begin{tabular}[c]{@{}c@{}}Post-test, \%\\mean (s.d.)\end{tabular} & \begin{tabular}[c]{@{}c@{}}Gain, pp\\mean (s.d.)\end{tabular} & Cohen's \textit{d} & \textit{P} & Statistic \\
\midrule
\multirow{2}{*}{1} & \multirow{2}{*}{102} & \multirow{2}{*}{24 (12)} & \multirow{2}{*}{0.89 [0.71--1.08]} & Immediate & 32 (15) & \cellcolor{gainlow}8 (15) & 0.53 & $<0.001$ & $W = 976.50$ \\
 &  &  &  & Delayed & 31 (14) & \cellcolor{gainlow}6 (16) & 0.38 & $<0.001$ & $t = -4.00$ \\
\addlinespace
\multirow{2}{*}{2} & \multirow{2}{*}{112} & \multirow{2}{*}{24 (13)} & \multirow{2}{*}{3.14 [2.58--3.60]} & Immediate & 64 (24) & \cellcolor{gainmid}39 (24) & 1.63 & $<0.001$ & $W = 53.00$ \\
 &  &  &  & Delayed & 39 (16) & \cellcolor{gainlow}15 (20) & 0.75 & $<0.001$ & $t = -8.14$ \\
\addlinespace
\multirow{2}{*}{3} & \multirow{2}{*}{109} & \multirow{2}{*}{24 (12)} & \multirow{2}{*}{11.11 [8.88--12.31]} & Immediate & 69 (25) & \cellcolor{gainhigh}45 (23) & 1.96 & $<0.001$ & $t = -20.25$ \\
 &  &  &  & Delayed & 40 (19) & \cellcolor{gainmid}17 (20) & 0.85 & $<0.001$ & $t = -8.41$ \\
\addlinespace
\multirow{2}{*}{4} & \multirow{2}{*}{108} & \multirow{2}{*}{23 (13)} & \multirow{2}{*}{36.92 [34.75--42.24]} & Immediate & 74 (24) & \cellcolor{gainvhigh}51 (28) & 1.82 & $<0.001$ & $W = 125.00$ \\
 &  &  &  & Delayed & 47 (20) & \cellcolor{gainmid}24 (22) & 1.09 & $<0.001$ & $t = -11.20$ \\
\bottomrule
\end{tabular}
\vspace{-1mm}

\begin{flushleft}
\footnotesize
Scores are expressed as percentages of the maximum rubric score. Pre-test, \textit{n}, and Duration are shared across the immediate and delayed rows within each condition (same participants, same baseline). Duration is the time-on-task for the assigned instructional condition, reported as the median with a bootstrap 95\% confidence interval (10,000 resamples); medians are reported because a small number of idle sessions produced extreme recorded durations. Gain is the within-condition change (post-test minus pre-test), in percentage points (pp). Gains were tested with two-sided paired \textit{t}-tests ($t$) or Wilcoxon signed-rank tests ($W$) where the normality assumption was violated (Shapiro--Wilk \textit{P} < 0.05); $\alpha = 0.05$. Paired \textit{t}-tests compared pre-test with post-test scores, so negative \textit{t} values correspond to positive gains. Effect sizes are Cohen's \textit{d}. Blue shading in the Gain column denotes larger within-condition gains, using a light sequential tint for visual emphasis only. s.d., standard deviation.
\vspace{-6mm}
\end{flushleft}
\end{table}

\textit{Gains scaled monotonically with the designed engagement level.} Mean immediate gains rose from $8$ percentage points in the brief passive reminder condition (Condition~1) to $39$ in scenario-based reading (Condition~2), $45$ in component selection (Condition~3), and $51$ in select-then-write (Condition~4): a more than sixfold increase from the lowest to the highest engagement-level condition (Table~\ref{tab:learning_gains}). One-way analysis of variance confirmed significant between-condition differences ($F(3, 427) = 62.4$, $p < .001$). Tukey HSD post-hoc tests indicated that Conditions~2--4 each produced significantly greater gains than Condition~1 (all $p < .001$), and that Condition~4 exceeded Condition~2 ($p < .01$); Conditions~4 and~3 did not differ significantly ($p = .08$). The ordering of mean gains predicted by ICAP held at the immediate post-test, with the strongest gains in the condition designed for the highest engagement (Condition~4). The same graded pattern persisted at the six-week delayed post-test (mean delayed gains $6$, $15$, $17$, and $24$ percentage points in Conditions~1--4; Table~\ref{tab:learning_gains}).

\textit{The graded pattern was not attributable to instructional time alone.} The four conditions differed substantially in time-on-task (median durations 0.89 to 36.92 minutes; Table~\ref{tab:learning_gains}), so we tested whether time, rather than the instructional design, accounted for the between-condition differences. Controlling for log-transformed time-on-task (winsorized at the within-condition 95th percentile to remove idle-session outliers), all between-condition differences in learning gains remained significant and essentially unchanged in magnitude. Time-on-task contributed a small additional amount to immediate gains ($b = 3.0$ percentage points per log-minute, $p = .023$) and did not predict delayed gains ($p = .83$): the durability of the gains, in particular, was unrelated to how long the instruction took.

\textit{Component-level gains were selective rather than uniform.} Gains were not spread evenly across the six rubric components (Supplementary Table S2): the gains of the brief reminder (Condition~1) were concentrated in the two components its message implies, guardrails ($+21.7$ pp) and tutoring protocol ($+22.3$ pp), with little movement elsewhere ($-5.2$ to $+7.8$ pp); the instructed conditions (Conditions~2--4) additionally gained $32$--$60$ pp on components with near-zero baselines ($\leq 3\%$), such as specifying the AI's role and the learner's level; and the two components students already largely supplied at baseline ($54$--$61\%$), difficulty identification and problem context, showed the smallest gains across conditions. At the delayed post-test, gains in specifying the learner's level and the tutoring protocol were above zero in all four conditions, whereas gains in the AI's role and guardrails declined substantially, most sharply after reading worked examples (Condition~2; AI's role from $48.2$ to $5.1$ pp), and gains in problem context and difficulty identification were no longer distinguishable from zero in any condition.

\subsection*{RQ2: Students sustained voluntary use and gained confidence, but academic performance did not differ between conditions}

To answer RQ2, we examined three outcomes: students' voluntary use of pedagogical prompting over the six weeks following the intervention, their confidence in prompting, and their performance on the course's final examination (Table~\ref{tab:secondary_outcomes}).

\begin{table}[!ht]
\centering
\small
\setlength{\tabcolsep}{5pt}
\caption{\textbf{Secondary outcomes by instructional condition.}}
\vspace{-2mm}
\label{tab:secondary_outcomes}
\begin{tabular}{lccccccc}
\toprule
 & Condition 1 & Condition 2 & Condition 3 & Condition 4 & $F$ & $p$ & $\eta^2$ \\
\midrule
Voluntary use$^{a}$ & 3.83 (0.83)$^{b}$ & 3.70 (1.00) & 3.50 (0.93)$^{b}$ & 3.77 (0.78) & 2.88 & \textbf{0.036} & 0.020 \\
Confidence gain$^{a,c}$ & 0.17 (0.85)$^{d}$ & 0.60 (1.00)$^{d}$ & 0.69 (0.99)$^{d}$ & 0.64 (0.95)$^{d}$ & 6.64 & $\textbf{<0.001}$ & 0.045 \\
Final exam score, \%$^{e,f}$ & 67.2 (16.8) & 70.8 (17.2) & 69.1 (17.6) & 71.0 (16.8) & 1.12 & 0.339 & 0.008 \\
\bottomrule
\end{tabular}
\vspace{-1mm}

\begin{flushleft}
\footnotesize
Values are mean (s.d.). Between-condition tests are one-way analyses of variance ($F(3,427)$, $n = 102$, $112$, $109$, $108$); effect sizes are $\eta^2$. $^{a}$Self-reported use of pedagogical prompting between the immediate and delayed post-tests, 1--5 scale. $^{b}$Condition~1 $>$ Condition~3 (Tukey HSD, $p = 0.032$); no other comparison significant. $^{c}$Change from the retrospective pre rating to the post rating, 1--5 scale. $^{d}$Conditions~2--4 each $>$ Condition~1 (Tukey HSD, all $p < 0.01$); no differences among Conditions~2--4. $^{e}$Final examination, \% of maximum score. $^{f}$Equivalence analysis: 90\% confidence intervals for all pairwise between-condition differences fell within $\pm 7.6$ percentage points; see text. s.d., standard deviation.
\vspace{-4mm}
\end{flushleft}
\end{table}

Self-reported \textit{voluntary use} of pedagogical prompting exceeded ``once per week'' in every condition ($M = 3.50$--$3.83$ on a 5-point scale). A between-condition analysis of variance reached significance ($F(3, 427) = 2.88$, $p = .036$, $\eta^2 = .020$); the only significant pairwise difference was that Condition~1 reported higher use than Condition~3 (Tukey HSD, $p = .032$). Moreover, responses to the open-ended item in delayed post-survey offered suggestive evidence of use beyond the course: without being asked, six students across all four conditions described applying learning-oriented prompting outside CSC108, four of them in a mathematics course (e.g., ``attaching previous exam reviews for my math class and asking it to provide me a similar but alternative mock exam, where i put my own answers to be graded'').

Gains in self-reported \textit{confidence} (post minus retrospective pre rating) were significantly larger in Conditions~2--4 ($M = 0.60$--$0.69$) than in Condition~1 ($M = 0.17$; $F(3, 427) = 6.64$, $p < .001$, $\eta^2 = .045$; Tukey HSD, all $p < .01$), with no differences among Conditions~2--4.

\textit{Final exam} scores did not differ significantly between conditions ($F(3, 427) = 1.12$, $p = .339$, $\eta^2 = .008$; means $67.2\%$, $70.8\%$, $69.1\%$, and $71.0\%$ in Conditions~1--4). An equivalence analysis contextualizes this null result: 90\% confidence intervals for all pairwise between-condition differences fell within $\pm 7.6$ percentage points, indicating that the data are inconsistent with moderate or large differences in examination performance (approximately $|d| \geq 0.45$), although equivalence within a stricter $\pm 5$-percentage-point margin could not be established for most comparisons. Additionally, at the individual level, confidence gains were positively correlated with prompting skill gains (Spearman $\rho = 0.31$, $p < .001$, $n = 431$). Confidence gains were, however, unrelated to final examination scores after controlling for condition, prompting skill gains, and pre-test scores ($b = -0.30$, $p = .74$), with no difference in this association across conditions (interaction $F(3, 422) = 1.24$, $p = .29$), and students whose confidence rose more than their prompting skill did not score lower on the examination ($\rho = -0.05$, $p = .34$).

\vspace{-1mm}

\subsection*{RQ3: Limited evidence of moderation across learner characteristics, with one exploratory gender interaction}

To examine potential heterogeneity in intervention effects, we tested moderation by baseline mindset, need for cognition, self-perceived programming ability (concept understanding and self-efficacy), gender, international student status, and program of study (Table \ref{tab:moderation}). We found little statistically reliable evidence that immediate or delayed gains varied across these measured characteristics. None of the four dispositional or self-perception variables significantly moderated gains, and no reliable moderation was detected by international status or program of study (Table \ref{tab:moderation}a).

One exploratory gender interaction was observed for immediate gains (Table \ref{tab:moderation}b). In the brief-reminder condition, women and non-binary students showed a numerically smaller mean gain than men (4.6 versus 9.3 percentage points), although this difference was imprecise and not statistically significant ($b=-4.7$, $p=.376$). Relative to the brief-reminder condition, the select-then-write condition showed a more positive immediate gain difference for women and non-binary students than for men ($b=17.1$ percentage points, $p=.024$). A similar directional pattern appeared in the scenario-based reading condition but was not statistically significant ($b=13.8$, $p=.061$). At the delayed assessment, the corresponding Condition~4 interaction remained directionally positive and approached statistical significance ($b=12.7$, $p=.059$). The immediate Condition~4 interaction was robust to permutation testing and baseline adjustment. However, it did not survive correction for multiple moderation tests, relied on the smaller subset of participants who reported gender, and only approached statistical significance at the delayed assessment. We therefore interpret this pattern as hypothesis-generating rather than confirmatory evidence and identify it as a priority for preregistered replication with adequately powered gender subgroups.

\begin{table}[!ht]
\centering
\small
\setlength{\tabcolsep}{5pt}
\caption{\textbf{Exploratory moderation analyses of learning gains by measured learner characteristics.}}
\label{tab:moderation}
\vspace{-2mm}
{\centering\textbf{Table 3a. }~Summary of exploratory moderation models.\par}\vspace{3pt}
\begin{tabular}{lcccc}
\toprule
 & \multicolumn{2}{c}{Immediate gain} & \multicolumn{2}{c}{Delayed gain} \\
\cmidrule(lr){2-3} \cmidrule(lr){4-5}
Learner characteristic & \begin{tabular}[c]{@{}c@{}}Main effect\\$\beta$ (s.e.)\end{tabular} & \begin{tabular}[c]{@{}c@{}}$\times$ Condition\\smallest $p$\end{tabular} & \begin{tabular}[c]{@{}c@{}}Main effect\\$\beta$ (s.e.)\end{tabular} & \begin{tabular}[c]{@{}c@{}}$\times$ Condition\\smallest $p$\end{tabular} \\
\midrule
Growth mindset & $-0.02$ (2.06), $p=.991$ & 0.231 & $-0.46$ (1.76), $p=.794$ & 0.446 \\
Need for cognition & $2.18$ (2.70), $p=.420$ & 0.118 & $-0.80$ (2.30), $p=.728$ & 0.359 \\
Self-perceived programming ability & $-1.26$ (1.49), $p=.397$ & 0.284 & $-0.29$ (1.28), $p=.819$ & 0.211 \\
Women and non-binary students (vs men)$^{a}$ & $-4.70$ (5.30), $p=.376$ & \cellcolor{gainmid}\textbf{0.024}$^{b}$ & $-1.23$ (4.69), $p=.793$ & 0.059 \\
International student (vs domestic) & $1.38$ (5.45), $p=.801$ & 0.187 & $-5.44$ (4.66), $p=.244$ & 0.091 \\
CS specialist/major (vs other program) & $5.01$ (6.33), $p=.430$ & 0.200 & $1.33$ (5.40), $p=.806$ & 0.078 \\
\bottomrule
\end{tabular}

\vspace{8pt}
{\centering \textbf{Table 3b. }~Full coefficients of the gender interaction models underlying the shaded row in \textbf{Table 3a}.\par}\vspace{3pt}
\begin{tabular}{lcccc}
\toprule
 & \multicolumn{2}{c}{Immediate gain} & \multicolumn{2}{c}{Delayed gain} \\
\cmidrule(lr){2-3} \cmidrule(lr){4-5}
Model term & $b$ (s.e.) & $p$ & $b$ (s.e.) & $p$ \\
\midrule
Intercept (Condition~1, men) & 9.33 (3.14) & 0.003 & 6.94 (2.78) & 0.013 \\
Condition~2 (vs Condition~1) & 28.07 (4.25) & $<0.001$ & 4.72 (3.76) & 0.211 \\
Condition~3 (vs Condition~1) & 36.27 (4.25) & $<0.001$ & 8.01 (3.76) & 0.034 \\
Condition~4 (vs Condition~1) & 37.98 (4.24) & $<0.001$ & 14.39 (3.75) & $<0.001$ \\
Women and non-binary (vs men) & $-4.70$ (5.30) & 0.376 & $-1.23$ (4.69) & 0.793 \\
Condition~2 $\times$ women and non-binary & 13.82 (7.34) & 0.061 & 11.10 (6.50) & 0.089 \\
Condition~3 $\times$ women and non-binary & 7.25 (7.98) & 0.364 & 7.81 (7.07) & 0.270 \\
Condition~4 $\times$ women and non-binary & \cellcolor{gainmid}\textbf{17.11 (7.53)} & \cellcolor{gainmid}\textbf{0.024} & 12.65 (6.67) & 0.059 \\
\bottomrule
\end{tabular}

\vspace{4pt}
\begin{flushleft}
\footnotesize
\vspace{-2mm}
(\textbf{Table 3a.})~Each row summarizes an exploratory ordinary-least-squares model predicting learning gain (in percentage points) from the learner characteristic, condition, and their interaction. Main-effect $\beta$ is the change in gain, in percentage points, per one-unit increase in the characteristic; for binary characteristics, $\beta$ is the difference relative to the reference group named in parentheses. $\times$ Condition reports the smallest unadjusted $p$ value among the three characteristic-by-condition interaction terms (Conditions~2--4 versus Condition~1). Blue shading marks uncorrected $p$ values below conventional thresholds: shading with bold, $p < .05$. No main effect reached statistical significance, and only the immediate Condition~4 gender interaction had an unadjusted $p<.05$. $^{a}$Gender was a skippable survey item ($n=328$ for this analysis). Women and non-binary students were combined because the non-binary subgroup was too small for a separate inferential comparison; a sensitivity analysis restricted to women versus men produced a similar directional pattern. $^{b}$The immediate Condition~4 interaction did not survive correction across the set of moderation models, and the corresponding delayed interaction only approached statistical significance. (\textbf{Table 3b.})~Full coefficient set of the exploratory gender models ($n=328$). The intercept is the mean gain of men in Condition~1; condition terms are condition effects among men; the gender term is the group difference within Condition~1; and interaction terms are difference-in-differences estimates relative to Condition~1. s.e., standard error.
\end{flushleft}
\end{table}

\section*{Discussion}

This study provides causal evidence that brief, single-session instruction can durably change how university students construct prompts when learning with GenAI. All four instructional approaches, from a two-sentence reminder to a multi-step constructive task, produced gains in learning-oriented prompting that persisted six weeks later, to the end of the semester, with the size of the gains scaling with the designed engagement level of the instruction (RQ1). The instruction's influence extended beyond the trained skill to self-reported sustained voluntary use and to students' confidence in prompting, although it did not differentiate broader academic performance between conditions (RQ2). We found little reliable evidence that gains varied across the measured learner characteristics, alongside one exploratory immediate gender interaction that warrants confirmatory replication (RQ3). As universities and instructors worldwide have adopted policies asking students to use GenAI ``for learning, not solutions''\cite{jin2025,sullivan2023}, these findings supply the missing link between such policies and classroom practice, in the form of a graded set of instructional options whose effects, and whose limits, are quantified rather than presumed.

\subsection*{What our instruction bridged: from GenAI policy to prompting practice}

\textbf{\textit{A mental-model-level message alone produces durable change \textbf{(RQ1)}.}} Condition~1 conveyed only a mental-model-level message, a two-sentence warning that using GenAI as an answer source can harm exam performance\cite{bastani2025}, yet it produced significant prompting gains that persisted six weeks later without reinforcement. This indicates that part of unproductive GenAI use stems from a default construal of the tool as an answer source rather than from missing technique; once that construal is updated, behavior shifts, consistent with minimal psychological interventions that change how people interpret a situation rather than what they know\cite{yeager2011,michie2011behaviour}. Condition~1's immediate gains concentrated in the guardrails and tutoring-protocol components that its two-sentence message implies, rather than spreading across the framework.

\textbf{\textit{Learning-oriented interaction with AI must be explicitly taught, and instruction designed for higher cognitive engagement produced larger, longer-lasting gains \textbf{(RQ1)}.}} Despite near-universal GenAI adoption\cite{hepi2026,stanfordhai2026}, students entered with pre-test scores averaging only 24\% of the rubric maximum, so everyday exposure to GenAI does not by itself produce learning-oriented prompting. Nor did the mental-model message alone close the gap: the brief reminder (Condition~1) yielded 8 percentage points immediate gain, whereas the conditions that explicitly taught and exercised the six components (Conditions~2--4) yielded 39 to 51, a gradient that held at six-week follow-up. That the gradient tracked the designed engagement levels is consistent with the ICAP framework\cite{chi2014,freeman2014} extending to an ill-defined competency without canonical solutions, although the conditions also differed in duration and in depth of exposure to the framework's content, so the gradient reflects the joint effect of these design features rather than a pure manipulation of engagement. We therefore read the four conditions not as isolated levels of a single psychological variable but as complete instructional options at realistic time costs, which is the comparison instructors face in practice. 

\begin{figure}[ht]
\centering
\includegraphics[width=\linewidth]{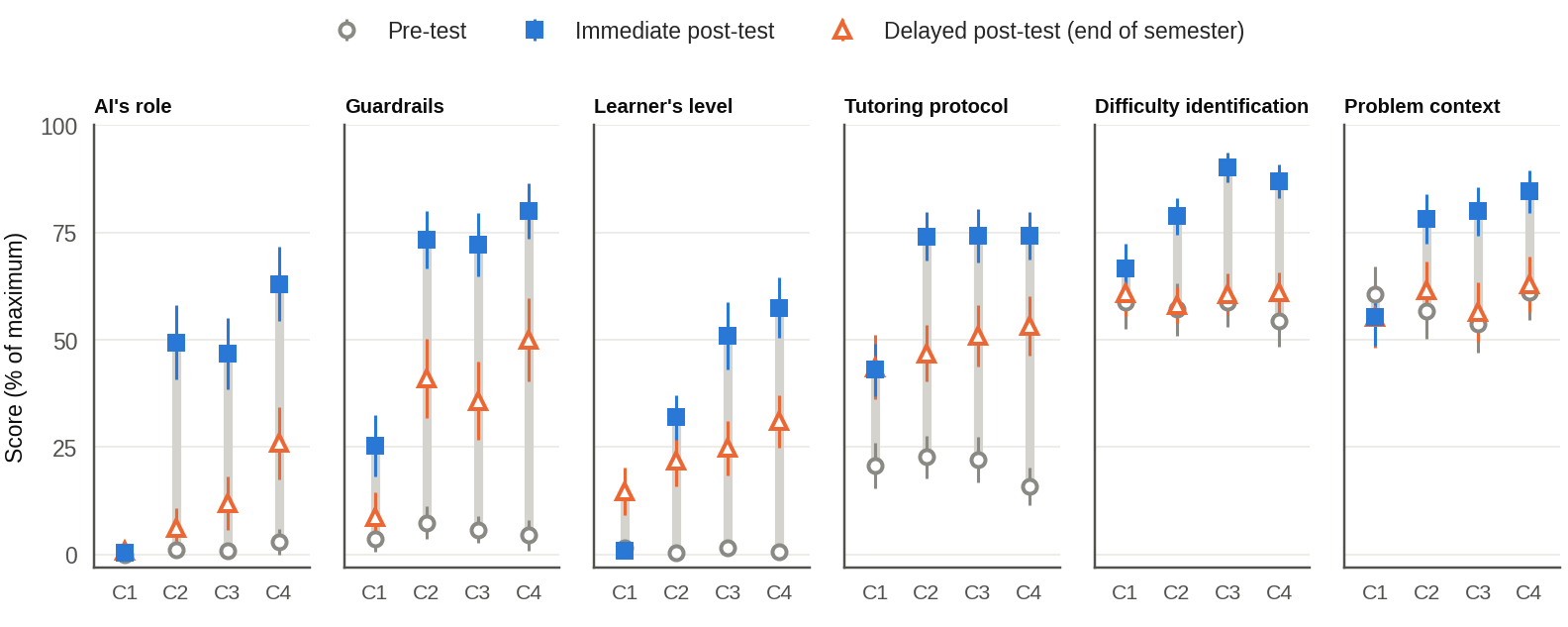}
\caption{Learning-oriented prompting by rubric component, condition, and time point. Points show mean scores as a percentage of the component maximum at pre-test, immediate post-test, and delayed post-test (end of semester), with 95\% confidence intervals; grey bars span the range across time points within each condition. Components are grouped, from left to right, into those rarely specified before instruction (the AI's role, guardrails, and the learner's level), the tutoring protocol, and those largely supplied before instruction (difficulty identification and problem context). Exact values and between-condition tests are reported in Supplementary Table S2. C1, brief reminder; C2, scenario-based reading; C3, component selection; C4, select-then-write.}
\label{fig:components}
\end{figure}

\textbf{\textit{The six components fall into three groups that call for different instructional choices (RQ1).}} The component-level results (Fig.~\ref{fig:components}; Supplementary Table S2) suggest three groups. First, most students already supplied difficulty identification and problem context before instruction, as pasting code and describing what went wrong is how students naturally seek help. Conditions~2--4 improved them beyond the brief reminder immediately, but the gains did not persist to the end of the semester, although the single delayed question limits this comparison. These components need little dedicated instruction. Second, the tutoring protocol improved substantially after the reminder alone; Conditions~2--4 added further immediate gains, but these faded to roughly the reminder's level by the end of the semester. Since specifying a pedagogical strategy may exceed what students can sustain after brief exposure, it may require more sustained instruction or be delegated to the system, with students choosing only the kind of support they need. Third, the AI's role, guardrails, and the learner's level were almost never specified before instruction. The reminder left the AI’s role unchanged, improved guardrails only briefly, and raised the learner’s level only at the end of the semester, whereas Conditions~2--4 produced the largest gains of all components, with the most active condition retaining the clearest advantage. These are the components where explicit instruction adds the most and that educators may most usefully target.

\textbf{\textit{Our instruction helps students distinguish the AI's fluency from their own understanding, avoiding the illusion of competence \textbf{(RQ2)}.}} When AI answers arrive polished and complete, learners readily mistake the AI's fluency for their own understanding (the ``illusion of competence''\cite{bjork1994,prather2024widening}). Consistent with our instruction fostering calibrated, rather than borrowed, confidence, confidence gains were largest in the instructed conditions, which were also the conditions with the largest objective skill gains, and individual confidence gains were positively correlated with individual skill gains (Spearman $\rho = 0.31$, $p < .001$, $n = 431$): self-belief moved together with demonstrated competence rather than outrunning it. Nor did this confidence appear to spill over into an illusion of competence about the course itself: confidence gains were unrelated to final examination scores, and students whose confidence rose more than their skill did not score lower. Although the association between confidence and skill gains is correlational, it suggests that such instruction can help students avoid the illusion of competence and build confidence in learning-oriented prompting that grows in step with their actual skill.

\textbf{\textit{Women and non-binary students may benefit more from explicit prompting instruction \textbf{(RQ3)}.}} Most measured learner characteristics did not reliably moderate immediate or delayed gains. One exploratory analysis suggested that the relative gender pattern was more positive in the select-then-write condition than in the brief-reminder condition at the immediate assessment. Specifically, women and non-binary students showed a larger relative immediate gain under Condition~4 than under Condition~1 when compared with men. This directional pattern is consistent with prior research suggesting that additional instructional structure can sometimes produce larger benefits for historically underrepresented learners in computing and STEM\cite{margolis2002unlocking,haak2011increased,theobald2020active}. One possible explanation is that explicit prompting instruction reduces reliance on uneven prior experience with GenAI. However, because the immediate interaction did not survive correction and only approached significance at the delayed assessment, this interpretation remains tentative. The finding therefore motivates confirmatory replication while encouraging educators to consider whether structured prompting instruction may create more inclusive conditions for participation in GenAI-enhanced learning.

\subsection*{What remains unbridged: from prompting skill to academic achievement}

\textbf{\textit{The skill gains did not translate into examination performance, and their durability argues against a too-weak intervention (RQ1).}} What the instruction did not do is produce a measurable change in academic performance. The durability of the skill gains argues against the simplest explanation, that the intervention was too weak or too transient to matter: all four conditions still showed significant gains at the delayed post-test, administered within the final examination itself.

\begin{figure}[!ht]
\centering
\includegraphics[width=.7\textwidth]{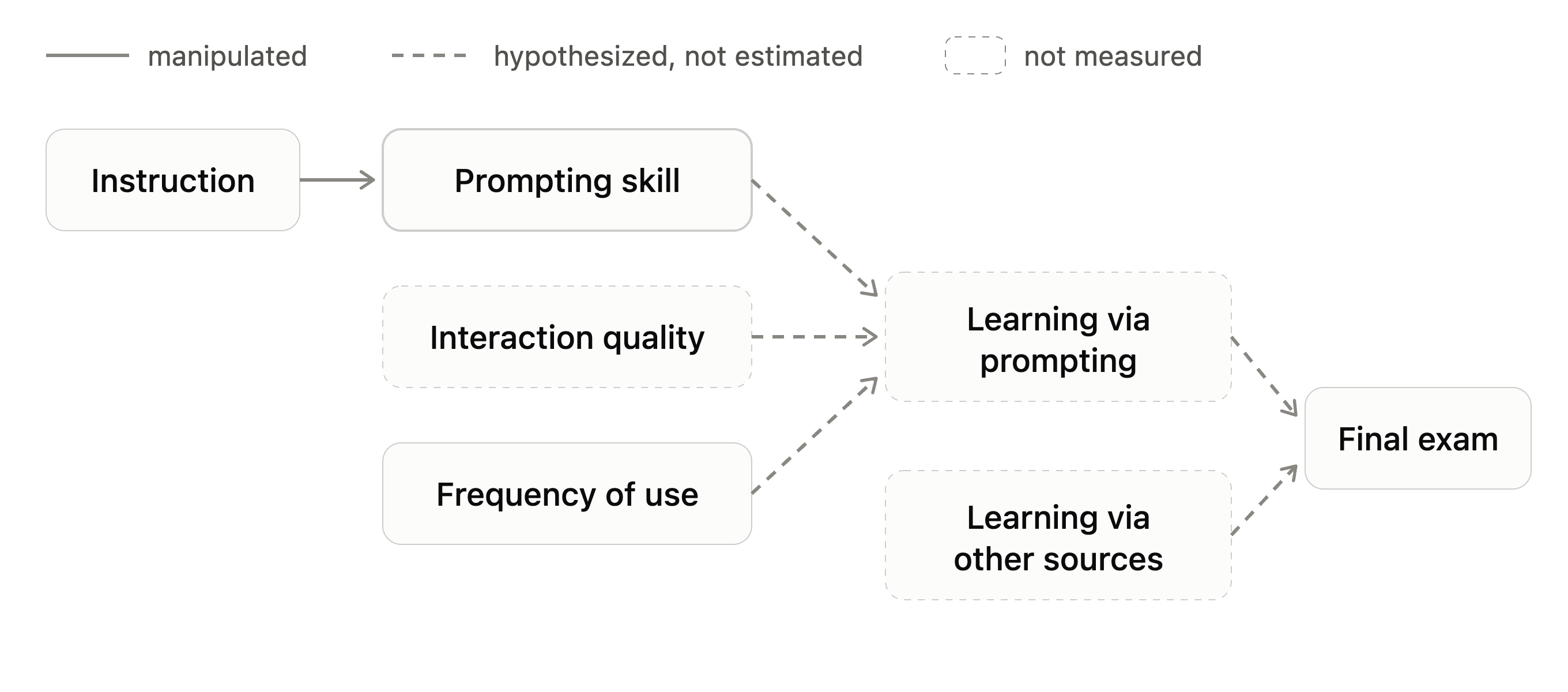}
\caption{Summary of the factors shaping how brief GenAI instruction translated policy into learning-oriented practice. The figure synthesizes the intervention components, observed outcomes, and boundary conditions discussed in the paper.}
\label{fig:pathway}
\end{figure}

\textbf{\textit{Two downstream antecedents likely gate the path from skill to achievement: frequency of use and interaction quality (RQ2).}} As Fig.~\ref{fig:pathway} sketches, we suspect the explanation lies downstream, in two antecedents that connect skill to achievement. The first is frequency: benefits can only accumulate with sufficient exposure, and prior work indicates that the relationship between GenAI use and achievement runs through how much students actually use the tools\cite{hepi2026,iqbal2025}; in our data, voluntary use, although high everywhere, did not follow the engagement gradient, so the conditions that built the most skill did not use it more. The second is interaction quality: a well-constructed prompt only opens a tutoring dialogue, and learning depends on what the student does inside it, with carefully structured tutoring interactions producing substantial gains\cite{kestin2025} and poorly structured ones impairing learning\cite{bastani2025,fan2025metacognitive}; this construct was not measured in the present study.

\textbf{\textit{Completing the bridge requires resolving the tension between frequency of use and prompting effort}, and measuring student-AI interaction naturalistically.} Our next steps therefore target both antecedents. Current agentic pedagogical applications\cite{wang2024,wang2026mas} that embed learning-oriented prompting into the AI system itself can make tutor-style interaction more convenient than composing prompts by hand\cite{kestin2025}, on the common assumption that reducing the effort of use will increase its frequency. Our results suggest that effort alone does not determine frequency. On the one hand, consistent with this assumption, students who received only the brief reminder (Condition~1) reported higher voluntary use than those in the more demanding active selection condition (Condition~3). On the other hand, the most demanding condition (Condition~4), in which students wrote each component themselves and revised it in response to feedback, sustained voluntary use as high as the brief reminder. Future work should therefore not judge an intervention's likely uptake by its effort alone, but test which effortful decisions must remain with students.

On the measurement of use side, naturalistic chat logs should replace self-report for dosage, and student turns in tutoring dialogues should be coded with multi-coder reliability to capture interaction quality; with both measured, mediation designs can test whether learning-oriented prompting causally contributes to course performance, completing the bridge this study began.

\subsection*{Which bridge to build: choosing an approach for one's students}
The four conditions form a graded menu for instructors (RQ1--RQ3). They range from under one minute to over half an hour of student time. Two findings simplify the choice. First, gains did not reliably depend on growth mindset, need for cognition, self-perceived programming ability, program of study, or international status. Instructors therefore need not tailor the approach to these characteristics, although the exploratory gender interaction warrants replication. Second, voluntary use exceeded once per week in every condition, and the most demanding condition did not reduce it. Instructors can thus choose primarily by available time and desired durability.

\textbf{\textit{Choosing Condition~1 for minimal class time.}} Instructors who can spare only a minute, for example when announcing a course AI policy, can use the brief reminder. Even this minimal prompt yielded measurable and durable gains: 8 percentage points immediately and 6 at the end of the semester. Voluntary use was also as high as in any other condition. The gains concentrated in guardrails and the tutoring protocol. This suggests that when prompted to self-regulate, learners' own notion of learning-oriented prompting centres on these two components: what the AI should not provide and how it should tutor. A policy-level reminder is therefore a useful floor rather than a symbolic gesture.

\textbf{\textit{Choosing Condition~2 for a low-cost introduction.}} Instructors who can spare a few minutes but cannot build interactive materials can use worked examples. Reading them takes about three minutes and requires only static text. It yielded a 39-point immediate gain, several times that of the reminder. It conveyed the AI-facing components (the AI's role, guardrails, and tutoring protocol) almost as well as active selection. Instructors should expect these gains to fade more by the end of the semester.

\textbf{\textit{Choosing Condition~3 for an efficient balance.}} Instructors who want strong immediate gains within a single class activity can use component selection. Per minute of instruction, the shorter conditions yielded more: about 9 points for the reminder and 12 for worked examples, versus 4 for component selection and 1.4 for select-then-write. However, the shorter conditions reached lower levels, and worked examples (Condition~2) improved specifying the learner's level less than component selection (Condition~3) did. Component selection reached 88\% of select-then-write's immediate gain (45 vs 51 points, not significantly different) in under a third of the time. Moving from component selection to select-then-write added only 6 points for 26 more minutes.

\textbf{\textit{Choosing Condition~4 for the strongest and most durable learning.}} Instructors who prioritize learning outcomes and can devote a longer session can use select-then-write. It takes about 37 minutes and requires automated feedback. In return, it produced the highest gains both immediately (51 points) and at the end of the semester (24 points, versus 17 for component selection). It was the only condition that still significantly exceeded all others in specifying the AI's role at the end of the semester. Despite being the most demanding, it did not discourage use: voluntary use was numerically higher than after component selection and comparable to that after the brief reminder.

\subsection*{Limitations and future work}
Several further caveats bound the present claims. Regarding measurement, our outcome rubric was derived from the same framework that organized instruction, so observed gains reflect adoption of that framework rather than framework-independent competence. Two facts bound this concern: no student saw the rubric at any test, and pre-test scores averaging 24\% of the rubric maximum indicate that the target behaviors are genuinely rare without instruction, so adopting an evidence-based framework is itself the instructional goal rather than an artifact of measurement. Relatedly, our tests measured whether students could write learning-oriented prompts when asked, not whether they did so in everyday coursework, where they may default to answer-seeking. The delayed post-test, completed without AI access six weeks later, shows the skill was retained, and the voluntary-use measure offers partial evidence of everyday application, though self-report shares the same social-desirability concern. Similarly, confidence before the intervention was rated retrospectively in the immediate post-survey, so confidence gains may partly reflect recall bias. Because the voluntary-use item asked about use in any context, it also cannot separate use within CSC108 from transfer to other courses. Future work should therefore measure more rigorously whether pedagogical prompting persists beyond the course and transfers to broader learning contexts, for example through students' naturalistic AI interaction logs across courses.

Regarding the experimental design, the conditions differed in duration and content exposure alongside their designed engagement levels, as discussed above, and we collected no direct manipulation check; isolating the contribution of cognitive engagement would require future designs that hold instructional time and content constant while varying the engagement demanded.

Regarding generalizability, the trial took place in a single introductory computer-science course at one institution, and the moderation analyses were not designed or powered as confirmatory tests of subgroup equivalence. Null interactions should therefore not be interpreted as evidence of established equity. Likewise, the exploratory immediate gender interaction did not survive correction across the set of moderation analyses and only approached statistical significance at the delayed assessment. Confirming whether explicit instruction differentially benefits learners with varied backgrounds will require preregistered replication with sufficiently large subgroup samples. Cross-disciplinary replication with framework-independent outcome measures is the natural complement to the application-pathway studies described above.

\section*{Author contributions}
R.X., R.Y., M.L., and J.S.\ led study design. R.Y.\ led system development. R.X., X.H., J.W., and H.K.\ contributed to data analysis and manuscript preparation. R.Y., M.L., and R.X.\ contributed to experiment deployment. H.K., M.L., and J.S.\ supervised the project. All authors contributed to manuscript preparation. All authors reviewed and approved the final manuscript.

\section*{Funding} 
This work was supported by Carnegie Mellon University through its Generative AI in Education seed grant (\$150{,}000; awarded June 2025). The funder had no role in study design, data collection and analysis, decision to publish, or preparation of the manuscript.

\section*{Competing Interests}
The authors declare no competing interests.

\section*{Approval for Human Experiments}
The study protocol was approved by the University of Toronto Research Ethics Board / Institutional Review Board (Protocol No. 00043701). All procedures were performed in accordance with relevant institutional guidelines and regulations. All participants provided informed consent before participation.

\section*{Data Availability}
Anonymized data and analysis code supporting the findings of this study are available from the corresponding author upon reasonable request and through the project repository, subject to institutional data-use agreements.

\bibliography{references}

\newif\ifappendsupplement
\appendsupplementtrue   
\ifappendsupplement
\clearpage
\setcounter{figure}{0}
\setcounter{table}{0}
\setcounter{section}{0}
\renewcommand{\thefigure}{S\arabic{figure}}
\renewcommand{\thetable}{S\arabic{table}}
\renewcommand{\figurename}{Supplementary Fig.}
\renewcommand{\tablename}{Supplementary Table}
\captionsetup{font=small,labelfont={bf,sf},justification=raggedright,singlelinecheck=false}
\begin{center}
{\LARGE\bfseries\sffamily Supplementary Information}\\[0.5em]
{\large Bridging GenAI policy and practice: Brief instruction improves learning-oriented prompting in an introductory programming course}
\end{center}
\vspace{1em}

\section*{Participant Flow}
Supplementary Fig.~\ref{fig:consort} shows the flow of participants through the trial, from allocation to the analytic sample.

\begin{figure}[htbp]
\centering
\includegraphics[width=\textwidth]{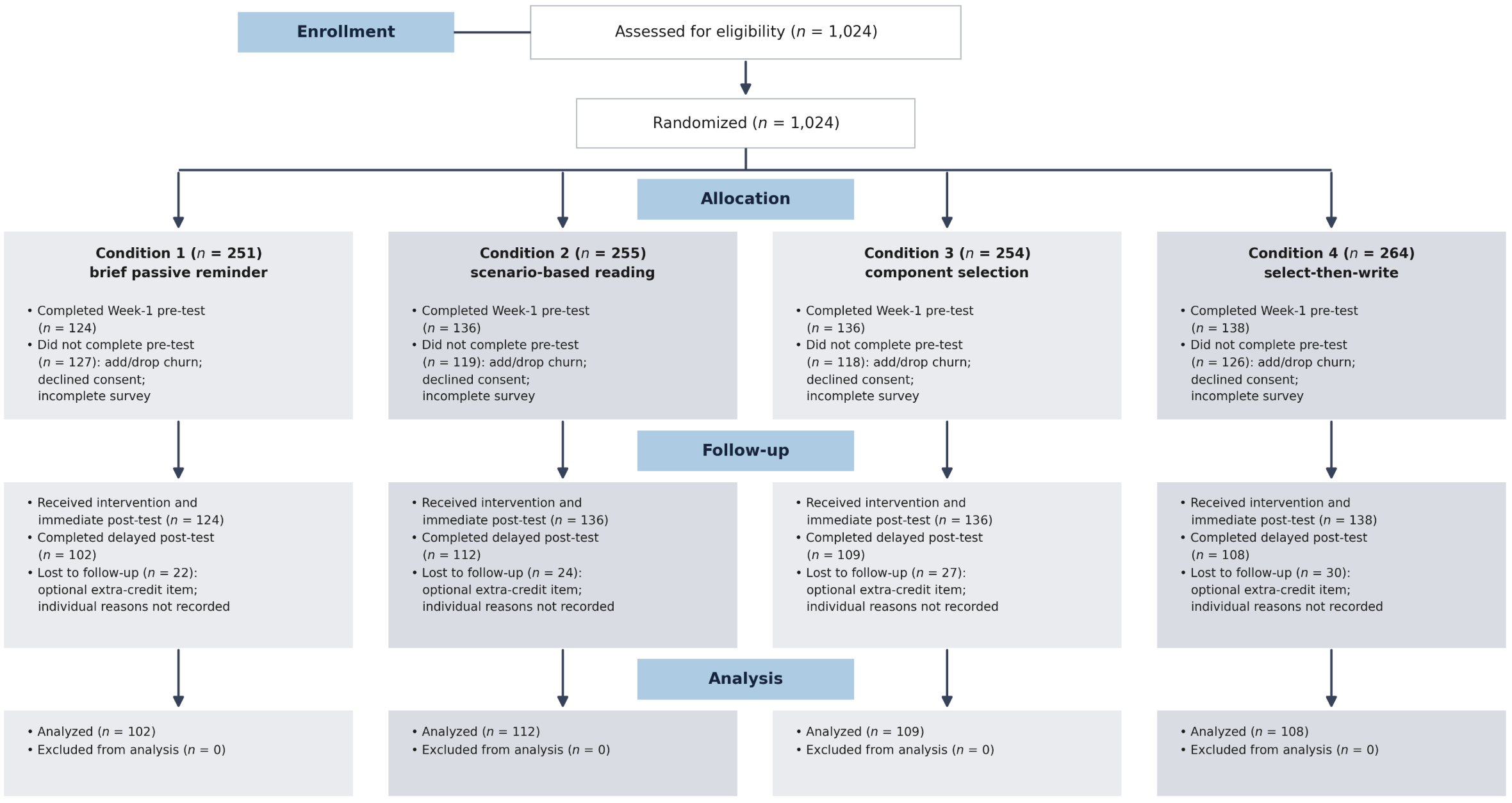}
\caption{CONSORT flow of participants through the trial. All 1,024 students enrolled in the mandatory course were assessed for eligibility and randomly assigned to one of four instructional conditions by a course-management script. The analytic sample comprises the 431 students who completed all four study phases. Attrition did not differ significantly between conditions at any stage or overall ($\chi^2(3) = 0.79$, $p = .852$), and students who did not complete the study did not differ from completers in baseline prompting scores (see Participants in the main text).}
\label{fig:consort}
\end{figure}

\section*{Condition Setup}
Students were randomly assigned to one of four online instructional conditions designed to increase the cognitive engagement required to practice learning-oriented prompting. All four conditions addressed the same six components of the prompting framework: the AI's role, the learner's level, problem context, difficulty identification, guardrails, and tutoring protocol. Figs.~\ref{fig:condition1_setup}--\ref{fig:condition4_setup} summarize the instructional activity in each condition.

Condition~1 (brief reminder) presented a two-sentence reminder to use GenAI as a tutor rather than as a shortcut; students read and acknowledged the reminder. Condition~2 (scenario-based worked examples) presented three illustrated programming-help scenarios together with example prompts, which students read without submitting a response. Condition~3 (component selection) asked students to work through the same scenarios and select the most appropriate option from three alternatives for each prompting component; a rule-based script checked the selections. Condition~4 (select-then-write) retained the selection activity and additionally asked students to write each component in their own words, with automated feedback to support revision when a component was missing or incomplete. Thus, Conditions~2--4 used the same scenarios and differed in the form of interaction: reading, selecting, and selecting plus generating, respectively.

The full text of the Condition~1 reminder is reproduced verbatim below.

\begin{quote}
\textbf{Use GenAI as a Tutor, Not a Shortcut}

\smallskip
Research from the Wharton School of the University of Pennsylvania found that non-tutoring use of GenAI (e.g., asking for final answers) was linked to about a 17\% drop in exam scores. Thus it is very important to make sure you are using Gen AI tools as a tutor, and ensure you actually learned after turning to GenAI for help in the rest of the semester.

\smallskip
This lab serves as a quick reminder of best practices when using GenAI tools as a tutor to help you learn, rather than as a shortcut to get answers. Click ``I Understand'' to conclude the lab.
\end{quote}

\begin{figure}[htbp]
\centering
\includegraphics[width=\textwidth,height=0.68\textheight,keepaspectratio]{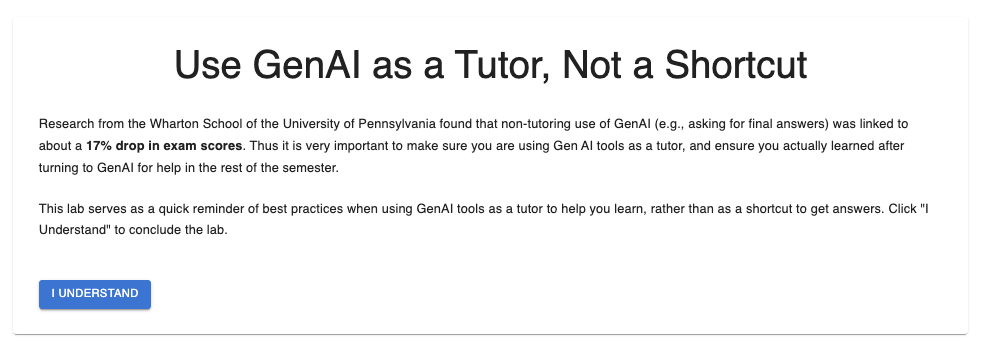}
\caption{Condition~1 interface: the brief reminder to use GenAI as a tutor rather than as a shortcut.}
\label{fig:condition1_setup}
\end{figure}

\begin{figure}[htbp]
\centering
\includegraphics[width=\textwidth,height=0.68\textheight,keepaspectratio]{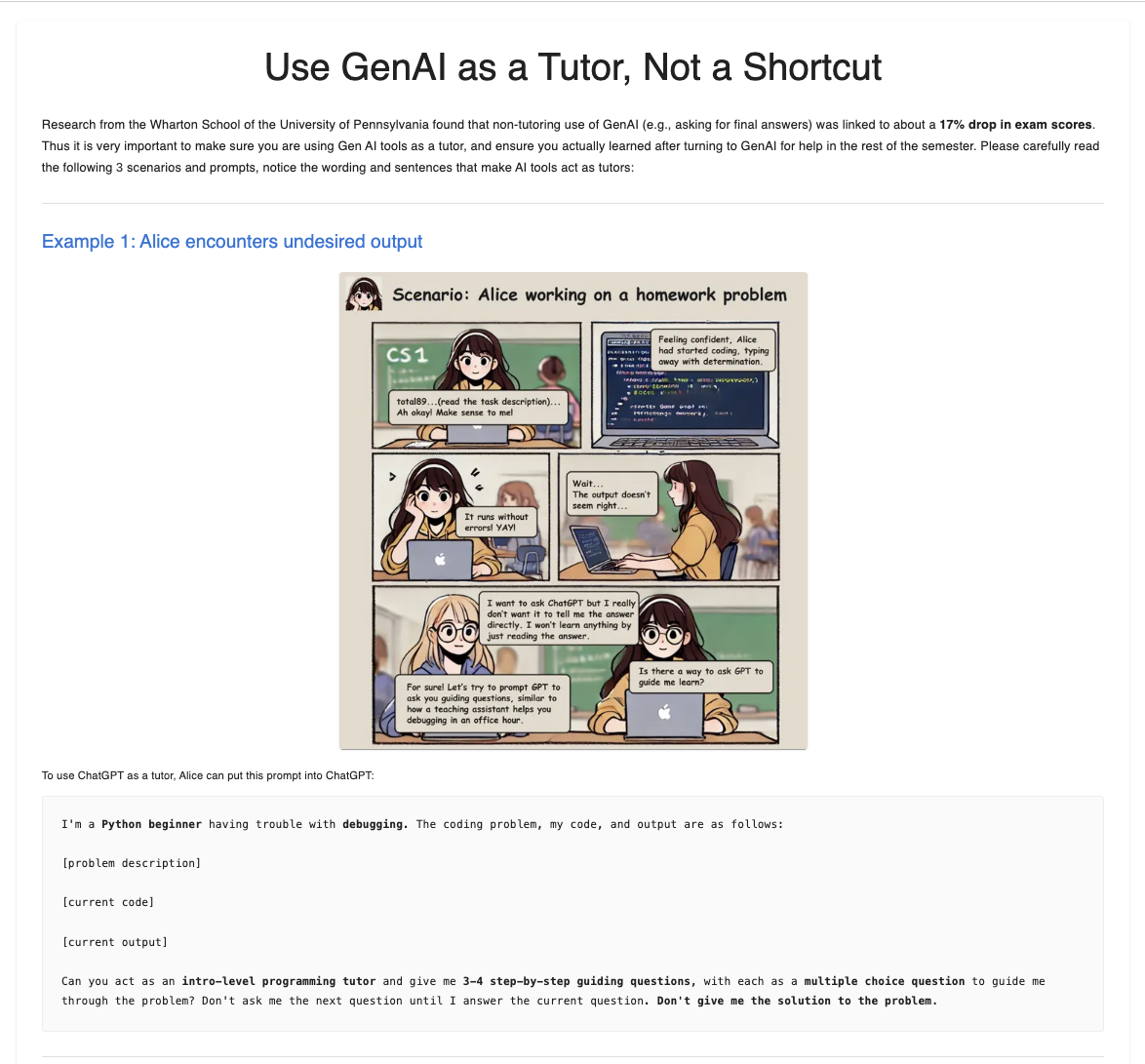}
\caption{Condition~2 interface: scenario-based worked examples in which students read illustrated programming-help scenarios and corresponding prompts.}
\label{fig:condition2_setup}
\end{figure}

\begin{figure}[htbp]
\centering
\includegraphics[width=\textwidth,height=0.68\textheight,keepaspectratio]{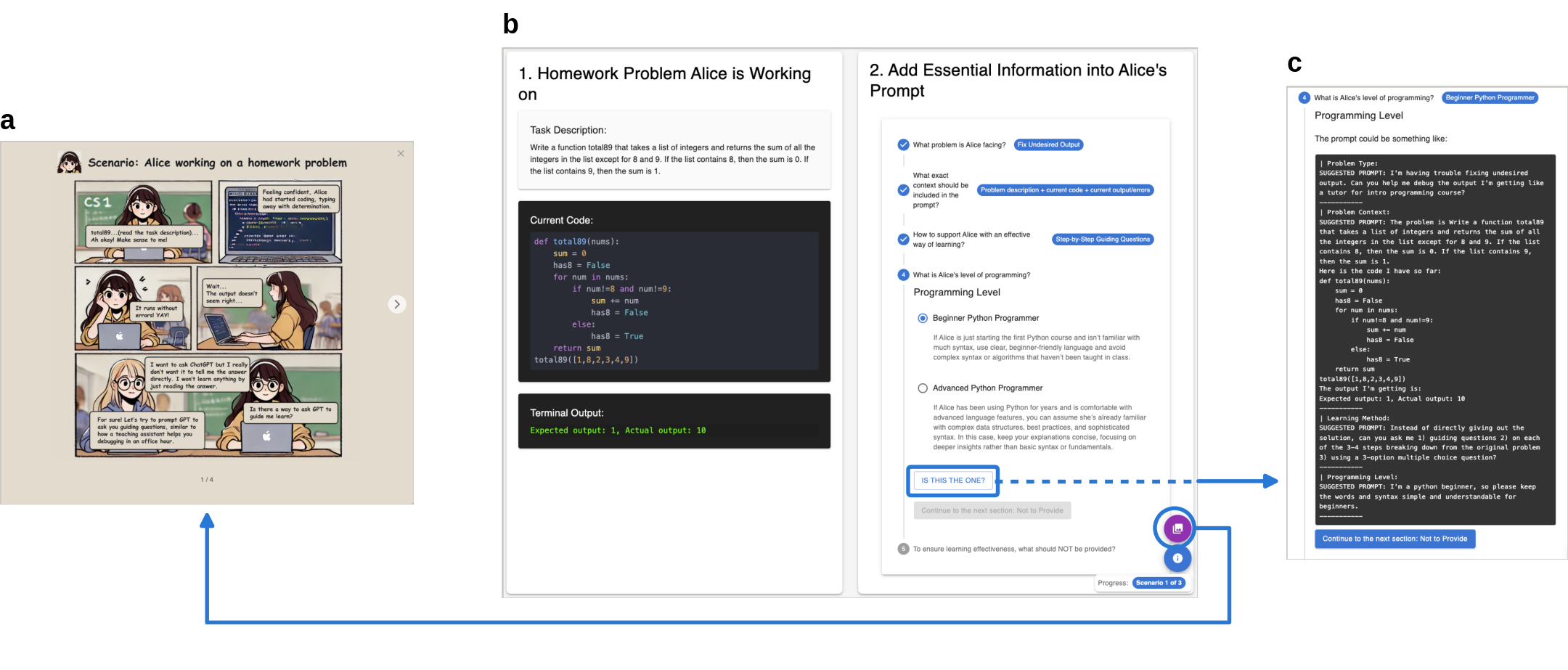}
\caption{Condition~3 interface: students select options for the prompting components in each scenario.}
\label{fig:condition3_setup}
\end{figure}

\begin{figure}[htbp]
\centering
\includegraphics[width=\textwidth,height=0.68\textheight,keepaspectratio]{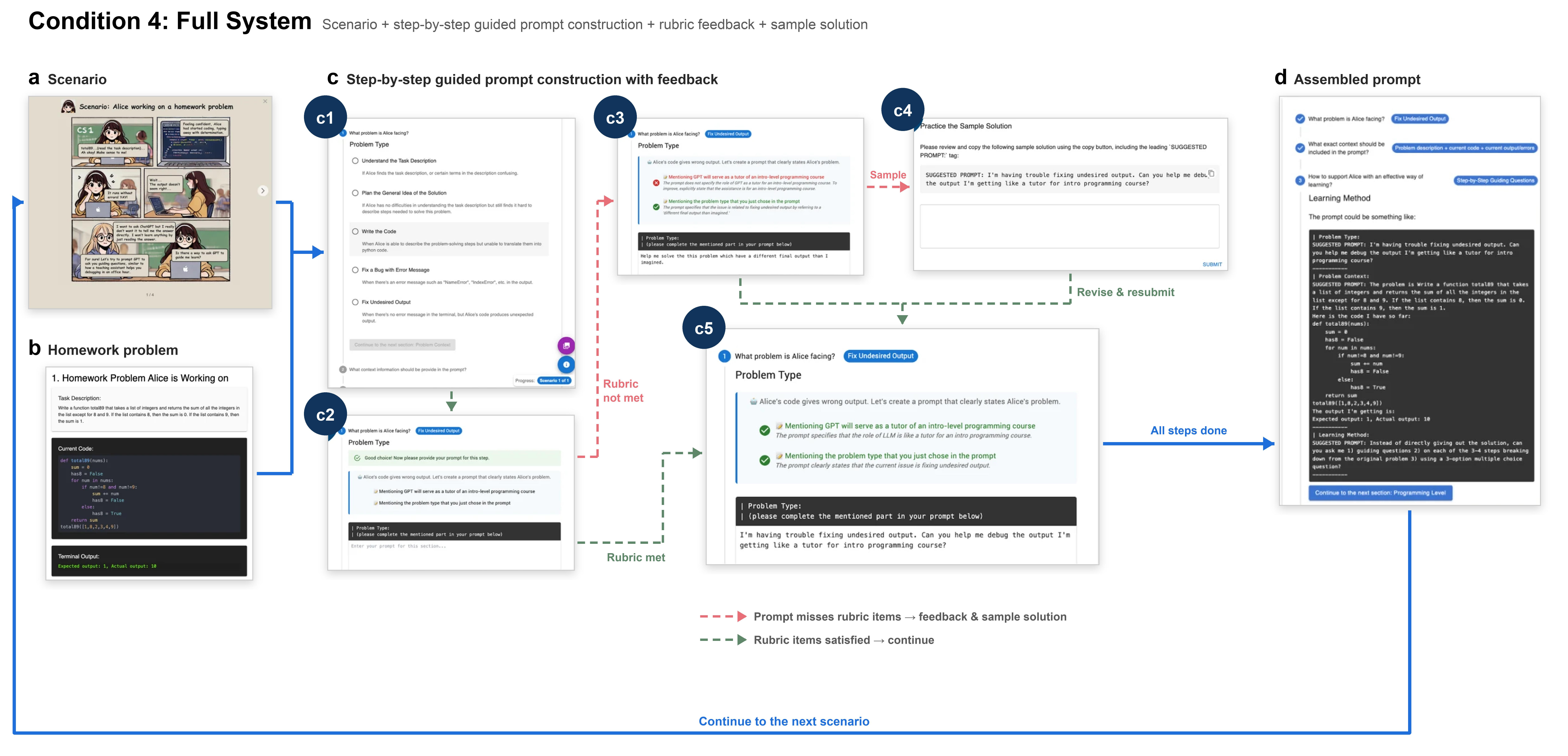}
\caption{Condition~4 interface: students select each prompting component, write it in their own words, and revise it using automated feedback.}
\label{fig:condition4_setup}
\end{figure}

\paragraph{Interactive demonstration of the four conditions.} The experimental system used in the study can be accessed at \url{https://llm-prompt-builder.vercel.app/}. Readers can experience each instructional condition by signing in with one of the following demonstration accounts, each of which is assigned to one condition and presents the same materials that participants in that condition received (student number \texttt{108108} for all accounts):
\begin{itemize}
\setlength{\itemsep}{0pt}
\item Condition 1 (brief reminder): UTORid \texttt{admin1}, student number \texttt{108108}
\item Condition 2 (scenario-based reading): UTORid \texttt{admin2}, student number \texttt{108108}
\item Condition 3 (component selection): UTORid \texttt{admin3}, student number \texttt{108108}
\item Condition 4 (select-then-write): UTORid \texttt{admin4}, student number \texttt{108108}
\end{itemize}

\section*{Pre- and Post-Tests}
The pre-test and immediate post-test each comprised three isomorphic programming-help scenarios in the format shown in Fig.~3a of the main text (task description, Alice's current code, and its output), on concepts taught in CSC108; scenarios differed between the two tests and from those used in the intervention. Before the first scenario, students read the following instructions:

\begin{quote}
You will be asked to write a prompt to ask ChatGPT to help student in a given scenario to learn what they currently struggle better. You are recommended to spend 1 minutes on getting a general understanding of the scenario, and then open the ChatGPT window, start a new chat session, spend 2 minutes to write the prompt and view GPT's response.
\end{quote}

\subsection*{Pre-test}
The pre-test was completed online at the end of the pre-survey in Week~1, with access to ChatGPT. For each scenario, students wrote a prompt in response to the item below, reproduced verbatim, and each prompt was scored with the six-component rubric.

\begin{quote}
Copy the prompt you used to ask ChatGPT here: \textit{(Scenarios 2 and 3: ``Start a new chat session to ask ChatGPT, then copy the prompt you used to ask ChatGPT here:'')}
\end{quote}

\subsection*{Immediate post-test}
The immediate post-test was completed online directly after the intervention, within the same take-home lab assignment, with access to generative AI chatbots. For each scenario, students wrote a prompt in response to the item below, reproduced verbatim, and each prompt was scored with the six-component rubric.

\begin{quote}
What prompt could Alice write to ask ChatGPT to guide her through the problem? Write the prompt below:

(Reminder: Your task here is NOT to solve this problem, rather, your task is to write a Generative AI prompt to have a system (like ChatGPT) teach the student in the scenario)
\end{quote}

\subsection*{Delayed post-test}
The delayed post-test was administered as an optional bonus question within the paper-and-pencil CSC108 final examination (Week~13), completed without access to AI. The text below is reproduced verbatim from the examination paper. The shared instructions preceded all three bonus questions; Question~1 constituted the delayed post-test, and Questions~2 and~3 constituted the delayed post-survey (see Pre- and Post-Survey).
 
\begin{quote}
\textbf{Bonus Question} \hfill (Optional, 3 pts)
 
\smallskip
\textbf{Note}: We \textbf{do not} expect you to be familiar with the technical content in this question. We would like to gain a better understanding of \textbf{your problem-solving skills}. By completing EVERY SECTION of this bonus, you will earn 3 marks. Leave anything blank, and you will forfeit the entire bonus.
 
\smallskip
\textbf{Prompt Writing}: Here is a CSC343 (Introduction to Databases) exam question that is outside the scope of CSC108. Imagine you are using ChatGPT to help you learn this new knowledge in advance of taking CSC343 with Professor Liut (you can refer to the question in the box below as \texttt{<question>} in your prompting):
 
\smallskip
\fbox{\begin{minipage}{0.92\linewidth}
\textit{Which SQL statement is used to add a foreign key constraint to an existing table?}
\begin{enumerate}
\renewcommand{\labelenumi}{\Alph{enumi}.}
\setlength{\itemsep}{2pt}
\item \texttt{ALTER TABLE childtable ADD CONSTRAINT fk\_name FOREIGN KEY (childcolumn) REFERENCES parenttable(parentcolumn);}
\item \texttt{ALTER TABLE childtable (childcolumn INTEGER, FOREIGN KEY (childcolumn) REFERENCES parenttable(parentcolumn));}
\item \texttt{UPDATE INTO childtable (childcolumn) VALUES (FOREIGN KEY (parentcolumn));}
\item \texttt{UPDATE TABLE childtable SET childcolumn = FOREIGN KEY (parentcolumn);}
\item \texttt{ALTER TABLE childtable SET FOREIGN KEY (childcolumn) REFERENCES parenttable(parentcolumn);}
\end{enumerate}
\end{minipage}}
 
\smallskip
\textbf{Requirements:} You must write your answer to all three questions on the next page in the provided space. Leaving any question blank or writing nonsense will result in a 0 for this bonus.
 
\medskip
\textbf{Question 1}: Write a prompt to ChatGPT, that \underline{you} would use to help yourself learn this content, so that you can answer a similar question on a future CSC343 exam.
\end{quote}
 
\section*{Pre- and Post-Survey}
\subsection*{Pre-survey}
The pre-survey was completed online in Week~1, before the pre-test. Only the items used in the present analyses are reproduced below, verbatim.

\paragraph{Demographic and background items.} Self-identified gender (``What is your self-identified gender? Please leave this question blank if you prefer not to answer.''; open text, coded as man, woman, or non-binary); intended programs of study (``What are your intended programs of study? (You may select more than one.)''; coded as computer science if any computer science major, specialist, or minor was selected); and student status (``Which of the following best describes you?''; Domestic student, International student).

\paragraph{Implicit theories of intelligence.} ``Please state your agreement with the following questions.'' Six-point scale (Strongly Disagree, Disagree, Somewhat Disagree, Somewhat Agree, Agree, Strongly Agree); items 1--4 were reverse-scored so that higher values indicate a stronger growth mindset.
\begin{enumerate}
\setlength{\itemsep}{0pt}
\item You have a certain amount of intelligence, and you can't really do much to change it.
\item Your intelligence is something about you that you can't change very much.
\item To be honest, you can't really change how intelligent you are.
\item You can learn new things, but you can't really change your basic intelligence.
\item No matter who you are, you can significantly change your intelligence.
\item You can always substantially change how intelligent you are.
\item You can change even your basic intelligence level considerably.
\item You can substantially increase your intelligence.
\end{enumerate}

\paragraph{Need for cognition.} ``Please state your agreement with the following questions.'' Items 3, 4, 5, 7, 8, 9, 12, 16, and 17 were reverse-scored.
\begin{enumerate}
\setlength{\itemsep}{0pt}
\item I would prefer complex to simple problems.
\item I like to have the responsibility of handling a situation that requires a lot of thinking.
\item Thinking is not my idea of fun.
\item I would rather do something that requires little thought than something that is sure to challenge my thinking abilities.
\item I try to anticipate and avoid situations where there is a likely chance I will have to think in depth about something.
\item I find satisfaction in deliberating hard and for long hours.
\item I only think as hard as I have to.
\item I prefer to think about small, daily projects to long-term ones.
\item I like tasks that require little thought once I've learned them.
\item The idea of relying on thought to make my way to the top appeals to me.
\item I really enjoy a task that involves coming up with new solutions to problems.
\item Learning new ways to think doesn't excite me very much.
\item I prefer my life to be filled with puzzles that I must solve.
\item The notion of thinking abstractly is appealing to me.
\item I would prefer a task that is intellectual, difficult, and important to one that is somewhat important but does not require much thought.
\item I feel relief rather than satisfaction after completing a task that required a lot of mental effort.
\item It's enough for me that something gets the job done; I don't care how or why it works.
\item I usually end up deliberating about issues even when they do not affect me personally.
\end{enumerate}

\paragraph{Self-perceived programming ability: concept understanding.} ``For each of the following programming concepts, select the statement that best describes your current level of understanding.'' Seven-level scale: A0, I am unfamiliar with this concept; A1, In general, I know what this concept means; A2, In general, I can recognize the syntax used by the programming language to represent this concept; B1, I can read and trace code that uses this concept with a few elements; B2, I can read and trace code that uses this concept with several different elements; C1, I can mostly recognize patterns of using this concept in different types of code; C2, I can easily recognize and explain the logic of code using this concept (abbreviated). An attention-check row (``select B2'') is omitted.
\begin{enumerate}
\setlength{\itemsep}{0pt}
\item Variables and assignment
\item Reading user input from and printing an output
\item Arithmetic, relational, and logic operators
\item Making decisions using conditional statements
\item Looping or iteration
\item Data collections such as arrays or lists
\item Methods and/or functions
\item Classes and/or objects
\end{enumerate}

\paragraph{Self-perceived programming ability: programming self-efficacy.} ``Please indicate how confident you are that you can perform each of the following tasks in Python. (1 = Not at all confident, 7 = Very confident)''
\begin{enumerate}
\setlength{\itemsep}{0pt}
\item Trace the execution of a Python program.
\item Trace a program with nested if/elif/else conditionals.
\item Trace a program that uses a list.
\item Trace a program that iterates over nested structures (e.g., a list of lists).
\item Trace a program that contains function or method calls with arguments and return values.
\item Trace a program that uses the same variable name in different scopes (e.g., local vs. global/enclosing).
\item Identify recursive patterns in a problem and write a recursive Python function that exploits this structure.
\item Trace the execution of a program that uses recursion.
\item Design a set of classes that leverages inheritance appropriately (e.g., base class + subclass in Python).
\item Explain the difference between a class and an instance (object) in Python.
\item Given a class hierarchy where multiple classes define a method with the same name, identify which method is called (i.e., method overriding / Python's MRO).
\item Write a working Python program even after hours of debugging new errors.
\item Correctly trace a failing program even if my prior attempts to trace similar programs didn't succeed.
\item Given a design and an incorrect Python program, locate the source of the error.
\item Use a systematic process (e.g., isolate--reproduce--hypothesize--test) to analyze a problem, design a solution, and implement it in code.
\item Design a small set of tests (e.g., assert statements or simple unittest cases) to assess a program's correctness.
\item Decompose a problem into a set of simpler subproblems (functions/modules).
\item Evaluate boolean expressions I encounter in Python (including truthiness and short-circuiting with and\slash or\slash not).
\item Construct a boolean expression that matches a plain-English condition (e.g., ``age is between 13 and 18 inclusive and has consent'').
\item Write a small program that defines and calls functions, using parameters, return values, and control flow (if\slash elif\slash else, for\slash while) correctly.
\end{enumerate}

\subsection*{Immediate post-survey}
The immediate post-survey was completed online directly after the immediate post-test. The two items used in the present analyses are reproduced verbatim below; both used a five-point agreement scale (Extremely Disagree, Somehow Disagree, Neither Agree or Disagree, Somehow Agree, Extremely Agree; coded 1--5), and confidence gain was computed as the first minus the second.

\begin{enumerate}
\setlength{\itemsep}{2pt}
\item Now I know how to effectively use LLM AI tools like ChatGPT to support my learning.
\item Before this experiment, I know how to effectively use LLM AI tools like ChatGPT to support my learning.
\end{enumerate}

\subsection*{Delayed post-survey}
The delayed post-survey comprised Questions~2 and~3 of the same examination bonus question, following the shared instructions reproduced under the delayed post-test (see Pre- and Post-Tests). Question~2 was the open-ended item analysed descriptively under RQ2, and Question~3 was the voluntary-use item (responses coded 1 = Never to 5 = Everyday). Both are reproduced verbatim below.
 
\begin{quote}
\textbf{Question 2}: Reflect on taking CSC108 this semester, now about Lab 7.5 (asynchronous lab) on Ethical Practices with AI. Write one prompt that \underline{you} personally have used that was helpful for your learning, and explain briefly why it was helpful to you.
 
\medskip
\textbf{Question 3}: Since completing Lab 7.5, how many times have \underline{you} prompted a generative AI tool (e.g., ChatGPT, Gemini) for pedagogical/learning purposes (e.g., asking to explain step-by-step, without revealing the solution). \textit{Circle \textbf{one} answer below.}
 
\smallskip
\begin{tabular}{@{}ccccc@{}}
A & B & C & D & E \\
Never & Once a Month & Once a Week & More than Once a Week & Everyday \\
\end{tabular}
\end{quote}

 \section*{Descriptive Data}
Supplementary Table~\ref{tab:descriptives} reports the descriptive statistics of all measures by condition at each time point, together with a test of baseline differences between conditions for each pre-intervention measure. Supplementary Table~\ref{tab:components} reports scores and gains for each rubric component by condition.
\begin{table}[ht]
\centering
\small
\caption{Descriptive statistics by condition at each time point. Values are mean (s.d.) for continuous measures and n (\%) of each condition's analytic sample for categorical measures. Baseline tests are reported for pre-intervention measures only: one-way analyses of variance for continuous measures and $\chi^2$ tests of independence for categorical measures, excluding students who did not report the characteristic (for gender, women and non-binary students were combined, as in the moderation analyses, because the non-binary subgroup was too small); no baseline measure differed significantly between conditions. Concept understanding and self-efficacy are the two self-perceived programming ability scales (see Pre-survey). Prompting scores are percentages of the maximum rubric score; the pre-test and immediate post-test are means across three questions, and the delayed post-test is a single question. Both confidence ratings were collected in the immediate post-survey, the pre-intervention (``retrospective pre'') rating retrospectively (``Before this experiment, \ldots''). Measures are based on up to 431 students, with minor missingness on individual pre-survey scales.}
\label{tab:descriptives}
\setlength{\tabcolsep}{3.5pt}
\begin{tabular}{@{}llccccc@{}}
\toprule
Measure & Time point & Condition 1 & Condition 2 & Condition 3 & Condition 4 & Baseline test \\
 & & ($n = 102$) & ($n = 112$) & ($n = 109$) & ($n = 108$) & \\
\midrule
\multicolumn{7}{@{}l}{\textit{Learning-oriented prompting score, \%}} \\
 & Pre-test            & 24 (12) & 24 (13) & 24 (12) & 23 (13) & $F(3, 427) = 0.16$, $p = .922$ \\
 & Immediate post-test & 32 (15) & 64 (24) & 69 (25) & 74 (24) & \\
 & Delayed post-test   & 31 (14) & 39 (16) & 40 (19) & 47 (20) & \\
\addlinespace
\multicolumn{7}{@{}l}{\textit{Confidence in prompting (1--5), immediate post-survey}} \\
  & Retrospective pre & 4.08 (0.89) & 3.90 (1.02) & 3.78 (0.82) & 3.86 (0.99) & $F(3, 427) = 1.90$, $p = .129$ \\
 & Post & 4.25 (0.72) & 4.50 (0.66) & 4.47 (0.65) & 4.50 (0.68) & \\
 & Gain                  & 0.17 (0.85) & 0.60 (1.00) & 0.69 (0.99) & 0.64 (0.95) & \\
\addlinespace
\multicolumn{7}{@{}l}{\textit{Baseline learner characteristics (pre-survey)}} \\
 & Growth mindset (1--6)        & 3.73 (1.13) & 3.58 (1.21) & 3.55 (1.05) & 3.50 (1.11) & $F(3, 426) = 0.77$, $p = .510$ \\
 & Need for cognition (1--7)    & 4.70 (0.86) & 4.71 (0.90) & 4.51 (0.85) & 4.54 (0.86) & $F(3, 426) = 1.59$, $p = .190$ \\
 & Concept understanding (1--7) & 4.54 (1.55) & 4.88 (1.38) & 4.60 (1.48) & 4.88 (1.46) & $F(3, 424) = 1.59$, $p = .192$ \\
 & Self-efficacy (1--7) & 4.43 (1.50) & 4.53 (1.44) & 4.34 (1.45) & 4.46 (1.44) & $F(3, 424) = 0.33$, $p = .803$ \\
\addlinespace
\multicolumn{7}{@{}l}{\textit{Gender, n (\%)}} \\
 & Woman & 26 (25) & 27 (24) & 18 (17) & 23 (21) & $\chi^2(3) = 2.98$, $p = .394$ \\
 & Man & 50 (49) & 60 (54) & 60 (55) & 61 (56) &  \\
 & Non-binary & 1 (1) & 1 (1) & 0 (0) & 1 (1) &  \\
 & Not reported & 25 (25) & 24 (21) & 31 (28) & 23 (21) &  \\
\addlinespace
\multicolumn{7}{@{}l}{\textit{Program of study, n (\%)}} \\
 & Computer Science & 72 (71) & 81 (72) & 78 (72) & 86 (80) & $\chi^2(3) = 2.74$, $p = .434$ \\
 & Other program & 17 (17) & 19 (17) & 20 (18) & 12 (11) &  \\
 & Not reported & 13 (13) & 12 (11) & 11 (10) & 10 (9) &  \\
\addlinespace
\multicolumn{7}{@{}l}{\textit{Student status, n (\%)}} \\
 & Domestic & 63 (62) & 63 (56) & 63 (58) & 66 (61) & $\chi^2(3) = 1.51$, $p = .679$ \\
 & International & 26 (25) & 37 (33) & 35 (32) & 32 (30) &  \\
 & Not reported & 13 (13) & 12 (11) & 11 (10) & 10 (9) &  \\
\addlinespace
\multicolumn{7}{@{}l}{\textit{Delayed outcomes}} \\
 & Voluntary use (1--5)      & 3.83 (0.83) & 3.70 (1.00) & 3.50 (0.93) & 3.77 (0.78) & \\
 & Final examination, \%     & 67.2 (16.8) & 70.8 (17.2) & 69.1 (17.6) & 71.0 (16.8) & \\
\bottomrule
\end{tabular}
\end{table}

\begin{table}[ht]
\centering
\footnotesize
\caption{Learning-oriented prompting by rubric component and condition.}
\label{tab:components}
\setlength{\tabcolsep}{3pt}
\begin{tabular}{@{}lccccc@{}}
\toprule
 & Condition 1 & Condition 2 & Condition 3 & Condition 4 & $F$(3, 427), $p$ \\
 & ($n = 102$) & ($n = 112$) & ($n = 109$) & ($n = 108$) & \\
\midrule
\multicolumn{6}{@{}l}{\textit{Pre-test, \%}} \\
\quad AI's role & 0 & 1 & 1 & 3 & 1.45, .228 \\
\quad Learner's level & 2 & 0 & 2 & 1 & 1.02, .383 \\
\quad Problem context & 61 & 57 & 54 & 61 & 1.06, .368 \\
\quad Difficulty identification & 59 & 57 & 59 & 54 & 0.46, .709 \\
\quad Guardrails & 4 & 7 & 6 & 4 & 0.95, .419 \\
\quad Tutoring protocol & 21 & 23 & 22 & 16 & 1.50, .214 \\
\addlinespace
\multicolumn{6}{@{}l}{\textit{Immediate gain, pp, mean [95\% CI]}} \\
\quad AI's role & 0.3 [$-$0.3, 0.9] & 48.2 [39.5, 56.9] & 45.9 [37.7, 54.1] & 60.0 [50.9, 69.1] & 44.61, $<$.001 \\
\quad Learner's level & $-$0.7 [$-$2.8, 1.4] & 31.5 [26.3, 36.7] & 49.2 [41.3, 57.1] & 56.8 [49.8, 63.8] & 68.01, $<$.001 \\
\quad Problem context & $-$5.2 [$-$12.6, 2.2] & 21.4 [13.2, 29.6] & 26.1 [18.6, 33.6] & 23.5 [15.3, 31.7] & 13.06, $<$.001 \\
\quad Difficulty identification & 7.8 [0.5, 15.1] & 21.7 [14.6, 28.8] & 31.3 [24.9, 37.7] & 32.6 [25.7, 39.5] & 10.34, $<$.001 \\
\quad Guardrails & 21.7 [14.6, 28.8] & 65.8 [58.0, 73.6] & 66.4 [59.0, 73.8] & 75.5 [68.1, 82.9] & 39.70, $<$.001 \\
\quad Tutoring protocol & 22.3 [15.3, 29.3] & 51.3 [44.0, 58.6] & 52.1 [44.7, 59.5] & 58.3 [51.1, 65.5] & 18.76, $<$.001 \\
\addlinespace
\multicolumn{6}{@{}l}{\textit{Delayed gain, pp, mean [95\% CI]}} \\
\quad AI's role & 1.0 [$-$0.9, 2.9] & 5.1 [0.1, 10.1] & 11.0 [5.0, 17.0] & 23.0 [14.4, 31.6] & 10.08, $<$.001 \\
\quad Learner's level & 13.1 [7.1, 19.1] & 21.4 [15.3, 27.5] & 23.1 [16.7, 29.5] & 30.4 [24.3, 36.5] & 5.09, .002 \\
\quad Problem context & $-$5.1 [$-$14.1, 3.9] & 4.8 [$-$4.2, 13.8] & 2.7 [$-$5.6, 11.0] & 2.0 [$-$6.6, 10.6] & 0.94, .421 \\
\quad Difficulty identification & 2.1 [$-$5.5, 9.7] & 1.0 [$-$6.5, 8.5] & 1.8 [$-$5.4, 9.0] & 6.8 [0.0, 13.6] & 0.52, .670 \\
\quad Guardrails & 5.2 [$-$0.7, 11.1] & 33.6 [23.5, 43.7] & 30.0 [20.8, 39.2] & 45.5 [35.5, 55.5] & 13.30, $<$.001 \\
\quad Tutoring protocol & 23.0 [14.4, 31.6] & 24.2 [16.6, 31.8] & 28.8 [19.7, 37.9] & 37.4 [29.8, 45.0] & 2.46, .062 \\
\bottomrule
\end{tabular}

\smallskip
\begin{minipage}{\linewidth}
\footnotesize
\textit{Values.} Pre-test values are mean percentages of the component maximum; gains are mean changes from the pre-test in percentage points (pp), with 95\% confidence intervals in brackets. All values are computed for the analytic sample.

\smallskip
\textit{Test items.} The pre-test and immediate post-test scores are means across three isomorphic questions on CSC108 content, whereas the delayed post-test is a single question on database foreign keys, a topic not covered in the course. Delayed gains for problem context and difficulty identification, which depend on the specific question, are therefore less directly comparable with immediate gains.

\smallskip
\textit{Between-condition tests.} The last column reports one-way analyses of variance comparing pre-test scores and gains across conditions; pre-test component scores did not differ between conditions. Significant pairwise differences (Tukey HSD, $p < .05$) were as follows. Immediate gains: Conditions~2--4 each exceeded Condition~1 on every component; Conditions~3 and~4 each exceeded Condition~2 on the learner's level (both $p < .001$); and Condition~4 exceeded Condition~3 on the AI's role ($p = .047$). Delayed gains: Condition~4 exceeded Conditions~1, 2, and~3 on the AI's role ($p < .001$, $p < .001$, and $p = .025$); Condition~4 exceeded Condition~1 on the learner's level ($p = .001$); and Conditions~2--4 each exceeded Condition~1 on guardrails (all $p \le .001$). All omnibus tests of gains with $p < .01$ remained significant after Holm correction across the 12 tests.
\end{minipage}
\end{table}

\section*{Scoring Rubric}
Each prompt was scored on the six components in Supplementary Table~\ref{tab:rubric}, each on a 0/0.5/1 scale; the total score (maximum 6) was expressed as a percentage of the maximum. No prompt received 0.5 on the AI's role or guardrails. The rubric was never shown to students. Examples are excerpts of student-written prompts from the pre-test and immediate post-test, reproduced verbatim; bracketed placeholders were written by students.

\begin{table}[ht]
\centering
\footnotesize
\caption{Scoring rubric for learning-oriented prompts, with example student prompts.}
\label{tab:rubric}
\renewcommand{\arraystretch}{1.2}
\begin{tabular}{@{}>{\raggedright\arraybackslash}p{0.13\linewidth}c>{\raggedright\arraybackslash}p{0.36\linewidth}>{\raggedright\arraybackslash}p{0.39\linewidth}@{}}
\toprule
Component & Score & Criterion & Example \\
\midrule
AI's role & 1 & The AI's role as a tutor, instructor, or teaching assistant is mentioned. & ``Can you act like my tutor and explain the task step by step\ldots'' \\
 & 0.5 & A role is assigned to the AI, but it is not a tutoring role (e.g., an expert programmer or a code reviewer). & \textit{Not observed in the data.} \\
 & 0 & Not mentioned. & ``Tell me what Alice did wrong'' \\
\addlinespace
Learner's level & 1 & The learner's beginner level is mentioned and specified, so that syntax and explanations stay simple. & ``I am beginner student in computer science\ldots\ as I am a beginner make the work simple and easy to understand'' \\
 & 0.5 & ``Beginner'' is mentioned without further specification. & ``I am a beginner python programmer can you guide me on spots of improvement without giving me the final answer?'' \\
 & 0 & Not mentioned. & ``What's wrong in this python code'' \\
\addlinespace
Problem context & 1 & The correct and complete problem context is provided (task, code, and output, if any). & ``my task is: [task instructions]. Here is my code and output as follows: [code], [output].'' \\
 & 0.5 & Part of the context is provided. & ``*paste the code above* edit the following code in order to reverse the number 12345'' \\
 & 0 & No context is provided. & ``Tell me what Alice did wrong'' \\
\addlinespace
Difficulty identification & 1 & The correct problem type is stated concisely (e.g., debugging, fixing wrong output, understanding the problem, code syntax). & ``what is the syntax for a for loop'' \\
 & 0.5 & The correct problem type is stated, but too long or too vaguely (e.g., ``I have a coding issue''). & ``do you fin d a mistake in this code?'' \\
 & 0 & Not mentioned. & ``explain and help me solve this problem'' \\
\addlinespace
Guardrails & 1 & The prompt asks the AI not to provide the direct solution. & ``Find the logic error but dont fix'' \\
 & 0.5 & The prompt limits the solution only partially or vaguely (e.g., asks for hints first without ruling out the full solution). & \textit{Not observed in the data.} \\
 & 0 & Not mentioned. & ``Tell me what Alice did wrong'' \\
\addlinespace
Tutoring protocol & 1 & A well-described learning strategy is specified (e.g., worked examples for a similar problem, explanation within the problem context, guiding questions in three steps, multiple-choice questions). & ``Can you give me 3-5 step-to-step guiding questions to help.'' \\
 & 0.5 & A learning strategy is mentioned but not well described. & ``Explain what I am supposed to do in simpler terms please'' \\
 & 0 & No learning strategy is mentioned. & ``Tell me what Alice did wrong'' \\
\bottomrule
\end{tabular}
\end{table}

\fi

\end{document}